\documentclass[aps,twocolumn,showpacs,preprintnumbers,amsmath,amssymb,nofootinbib,superscriptaddress,showkeys]{revtex4}

\usepackage{epsfig}
\usepackage{graphicx}

\begin{document}

\title{Renormalization of the Deuteron with One Pion Exchange}
  \author{M. Pav\'on Valderrama}\email{mpavon@ugr.es}
  \affiliation{Departamento de F\'{\i}sica Moderna, Universidad de
  Granada, E-18071 Granada, Spain.}  \author{E. Ruiz
  Arriola}\email{earriola@ugr.es} \affiliation{Departamento de
  F\'{\i}sica Moderna, Universidad de Granada, E-18071 Granada,
  Spain.}

\date{\today}

\begin{abstract} 
\rule{0ex}{3ex} We analyze the deuteron bound state through the One Pion
Exchange Potential. We pay attention to the short distance peculiar
singularity structure of the bound state wave functions in coordinate
space and the elimination of short distance ambiguities by selecting
the regular solution at the origin. We determine the so far elusive
amplitude of the converging exponential solutions at the origin.  All
bound state deuteron properties can then be uniquely deduced from the
deuteron binding energy, the pion-nucleon coupling constant and pion
mass. This generates correlations among
deuteron properties. Scattering phase shifts and low energy parameters
in the $^3S_1-{}^3D_1 $ channel are constructed by requiring
orthogonality of the positive energy states to the deuteron bound
state, yielding an energy independent combination of boundary
conditions. We also analyze from the viewpoint of short distance
boundary conditions the weak binding regime on the light of long
distance perturbation theory and discuss the approach to the chiral
limit.
\end{abstract}

\maketitle



\section{Introduction}

Pion dynamics plays a dominant role in the low energy structure of the
Nucleon-Nucleon interaction, and in particular in the description of
light nuclei like the deuteron~\cite{Ericson:1988gk}. The long distance part
of the interaction is given by one, two and higher pion exchanges and
the fact that the deuteron is a weakly bound state suggests that many
of its properties can indeed be explained in terms of these dynamical
degrees of freedom in a model independent way and regardless on the
less known short distance interaction. Glendenning and
Kramer~\cite{Glendenning62} in the early sixties recognized clear
correlations between several deuteron observables generated by
truncating the One Pion Exchange (OPE) potential at a distance
$R=0.4915 {\rm fm}$ and assuming a hard core inside. Tight constraints
on deuteron observables were established by Klarsfeld, Martorell and
Sprung~\cite{Klarsfeld:1980,Klarsfeld:1984es} by integrating the
deuteron wave function from infinity down to a cut-off radius using
the OPE potential and rigorous inequalities. An accurate determination
of the D/S asymptotic ratio was made by Ericson and
Rosa-Clot~\cite{Ericson:1981tn,Ericson:1982ei} based on assuming the
OPE correlation between the S and D wave functions and taking
realistic potential models to describe the S wave function.  (for a
review on these developments see e.g Ref.~\cite{Ericson:1985}). Friar
{\it et al}. use a multipole form factor~\cite{Friar84} whereas Ballot
{\it et al.}  used separate monopole forms factor for the central and
tensor part of the OPE potential mimicking the finite size of
nucleons~\cite{Ballot:1989jk,Ballot:1992sc}. Along a similar line of
investigation Sprung {\it et al.}  used a square well
potential~\cite{Martorell94} for the central component and a vanishing
potential for the tensor component.

Within the effective field theory (EFT) approach to nuclear physics
proposed by Weinberg~\cite{Weinberg:rz} the situation was revisited
from a somewhat different perspective since the OPE potential appears
as the lowest order of a perturbative hierarchy based on chiral
symmetry~\cite{Ordonez:1995rz} (for a review see e.g.
Ref.~\cite{Bedaque:2002mn}), and short distance ambiguities could be
eliminated by the renormalization program if the auxiliary regulator
is removed from the theory at the end of the calculation. This cut-off
independence should occur at {\it any level} of approximation, no
matter how many pions are exchanged. At long distances,
renormalization group methods suggest that one is close to an infrared
fixed point~\cite{Birse:1998dk}. The renormalization procedure can be
explicitly and analytically carried out within perturbation
theory~\cite{Kaplan:1998sz}. However, these nice features become a non
trivial numerical problem beyond perturbation theory motivating the
use of truncation cut-off schemes. The work of Ref.~\cite{Park:1998cu}
uses a gaussian cut-off in coordinate space to regulate the contact
delta interaction, Ref.~\cite{Frederico:1999ps} proposes the use of a
subtraction method in momentum space regulating the central part,
Ref.~\cite{Epelbaum:1999dj} uses a sharp momentum cut-off and in
Ref.~\cite{Phillips:1999am} it was proposed to use a finite short
distance cut-off, whereas Ref.~\cite{Entem:2001cg} puts exponentially
suppressed regulators in momentum space. It should be mentioned that
in all cases the corresponding coordinate/momentum space cut-off
parameter $a$/$\Lambda$ is uncomfortably large/small from the
viewpoint of renormalization theory. Typically, one has $a \sim 1.4
{\rm fm}$ (see e.g. Ref.~\cite{Phillips:1999am}) and $\Lambda=600 {\rm
MeV}$~\cite{Park:1998cu} respectively. So, it is not obvious that
according to the basic principles of EFT the short distance
ambiguities are, as one might expect, indeed under control. Moreover,
the existence of a well behaved finite renormalized limit is never
guaranteed {\it a priori} and one relies mainly on numerics. Actually,
the fact that the results on deuteron observables look rather similar,
regardless on the particular way how the potential is modeled at short
distances, proves that the long distance pion dynamics dominates the
physics confirming the findings of Glendenning and
Kramer~\cite{Glendenning62} more than 40 years ago, but does not
resolve the mathematical problem whether the OPE potential can make
unambiguous predictions regardless of any short distance physical
scale.

The OPE potential is local in coordinate space where the problem is
naturally formulated by the standard Schr\"odinger
framework. Moreover, it is singular at the origin and giving boundary
conditions at that point is not a well defined procedure for uniquely
determining the solution~\cite{Case:1950} (for a comprehensive review
in the one channel case see e.g.  Ref.~\cite{Frank:1971} ). There is
the added difficulty that we have two coupled second order
differential equations. In the deuteron channel one has four
independent solutions, which according to their singularity structure
correspond to either two regular and two irregular solutions at
infinity or three regular and one irregular solution at the origin.
The normalizability condition of the deuteron wave functions
eliminates all constants for a given deuteron binding energy, which
instead of being predicted has to be treated as an independent
parameter. One of the advantages of the coordinate space treatment of
renormalization is that it can directly be extended to other singular
cases such as the Two Pion Exchange (TPE) potential~\cite{Pavon_TPE}
which is also finite everywhere except the origin. In contrast,
momentum space treatments require an extra regularization of the
potential besides the standard cut-off regularization of the
Lippmann-Schwinger equation.

The authors of Ref.~\cite{Martorell94} found a discrete sequence of
equivalent short distance cut-off radii having almost the same
deuteron properties. In their analysis of the problem one regular
solution at the origin with a converging exponential behavior, $\exp
(-4(2 R/r)^{\frac12})$ with $ R \sim 1 {\rm fm}$, was discarded on
numerical grounds. The same result was also implicitly used in
Ref.~\cite{Beane:2001bc} and large $N_c$ arguments in favor of it were
raised in Ref.~\cite{Beane:2002ab}.  This extra condition would
actually {\it predict } the deuteron binding energy from the OPE
potential. As we will show in this paper, the converging exponential
is non-vanishing although rather elusive because its contribution to
the deuteron wave function only becomes sizeable at relatively large
distances and accurate numerical work must be done to pin down its value
with a certain degree of confidence.

In the present work we show that there is no need to truncate the OPE
potential on a physical scale to produce unique and cut-off
independent predictions for deuteron properties and scattering
observables in terms of the OPE potential parameters and the deuteron
binding energy. These might then legitimately be called OPE model
independent predictions and paves the way for a systematic
investigation on the case where more pions are exchanged and other
effects are taken into account~\cite{Pavon_TPE}. After presenting the
basic notation in Sect.~\ref{sec:ope}, we discuss in
Sect.~\ref{sec:short0} the regular solutions at the origin and
establish that the limit when the regulator is removed is finite. For
numerical purposes it is useful to define {\it some} short distance
regulator as an auxiliary tool. In Sect.~\ref{sec:cut-off} we use six
different regulators based on boundary conditions of the wave function
and check for stability to high precision for all regulators.  This
procedure generates correlations among deuteron observables if the
deuteron binding energy is varied as a free and independent parameter
as we do in Sect.~\ref{sec:OPE-corr}. A particularly interesting
situation is provided by the weak binding limit, which can be taken
with fixed OPE potential parameters. In such a case the long distance
behavior should dominate and one might expect perturbative methods to
apply and be compared to the exact OPE calculation. The details of the
perturbative calculation are postponed to Appendix~\ref{sec:pert}
where we present a coordinate space version of the method, in
consonance with the exact treatment. A detailed comparison shows that
the perturbative argument is too naive and would only hold in the weak
coupling regime as well, due to the appearance of non-analytical
contributions in the $\pi NN $ coupling constant. Mathematically, we
show that it is not possible to go beyond first order since the
coefficients of the expansion diverge. Numerically, the disagreement
at first order is typically on the $30 \%$ level for physical values
of the OPE parameters at zero binding. After Ref.~\cite{Bulgac:1997ji}
the chiral limit in nuclear physics has attracted considerable
attention in recent
works~\cite{Beane:2002xf,Beane:2002vs,Epelbaum:2002gb} and also the
limit of heavy pions in connection with lattice QCD calculations,
where the pion mass is still far from its physical value. We study in
Sect.~\ref{sec:m_dependence} the correlations among those observables
if the pion mass is varied away from its physical value by studying a
suitable extension of the Feynman-Helmann theorem. Another remarkable
property of the OPE potential which we deal with in
Sect.~\ref{sec:scattering} is that low energy parameters as well as
the scattering phase-shifts can be uniquely determined from the OPE
potential parameters and the deuteron binding, due to orthogonality
constraints of the bound state and scattering states. In
Sect.~\ref{sec:short} the determination of the non-vanishing
coefficient of the converging exponential at the origin is carried out
by a short distance expansion to eighth order of the OPE deuteron wave
functions. In Sect.~\ref{sec:concl} we come to the conclusions.

One of the surprising results in the OPE description of the deuteron
has to do with the small asymptotic ratio between the D and S waves, $
w(\infty) /u(\infty) = \eta = 0.0256$ coming from a large ratio at
short distances of order unity, $ w(0) / u(0) = 1/\sqrt{2} = 0.707
$. Although this feature is specific to the OPE potential it is
somewhat a bit outside the main topic of this work.  So we relegate
this issue to Appendix~\ref{sec:local_rot} where we show how this can
be easily understood if a local rotation of the deuteron wave
functions diagonalizing the coupled channel potential is carried
out. Obviously, such a transformation cannot simultaneously
diagonalize the kinetic terms, but the residual mixing is related to
the derivative of a local mixing angle which numerically turns
out to be a slowly varying function. Using this as a starting
approximation we can determine in a perturbative fashion the
asymptotic D/S ratio yielding the exact OPE value with a $1\%$
accuracy.

\section{Bound state equations and their solutions} 

\subsection{The OPE deuteron Equations}
\label{sec:ope} 

The Deuteron coupled channel $^3S_1 - {}^3D_1 $ set of equations read
\begin{eqnarray}
-u '' (r) + U_s (r) u (r) + U_{sd} (r) w (r) &=& -\gamma^2 u
 (r) \, ,\nonumber  \\ \\ -w '' (r) + U_{sd} (r) u (r) + \left[U_{d} (r) +
 \frac{6}{r^2} \right] w (r) &=& -\gamma^2 w (r) \, , \nonumber \\
\label{eq:sch_coupled} 
\end{eqnarray}
together with the asymptotic conditions at infinity  
\begin{eqnarray}
u (r) &\to & A_S e^{-\gamma r} \, , \nonumber \\ w (r) & \to & A_D
e^{-\gamma r} \left( 1 + \frac{3}{\gamma r} + \frac{3}{(\gamma r)^2}
\right) \, ,
\label{eq:bcinfty_coupled} 
\end{eqnarray}
where $ \gamma = \sqrt{M B} $ is the deuteron wave number, $A_S$ is
the normalization factor and the asymptotic D/S ratio parameter is
defined by $\eta=A_D/A_S$.  The $^3S_1-{}^3D_1 $ coupled channel
potential is given by
\begin{eqnarray}
U_s = U_c \, , \qquad U_{sd} = 2 \sqrt{2} U_T \, , \qquad U_d = U_C -
2 U_T \, ,
\end{eqnarray} 
where the OPE reduced potential ($U=2 \mu V $) is given for $r > 0 $
by
\begin{eqnarray}
U_C &=& -\frac{m M g_A^2 }{16 \pi f_\pi^2 } \frac{e^{-m r 
}}{r} \, \\ U_T &=& -\frac{m^2 M g_A^2 }{16 \pi f_\pi^2 }
\frac{e^{-m r }}{r} \left( 1 + \frac3{m r}+ \frac3{(m
r)^2} \right) \, , 
\end{eqnarray} 
where $m$ is the pion mass, $M=2 \mu_{np} = 2 M_n M_p /(M_n+M_p) $
twice the reduced proton-neutron mass, $g_A$ the axial nucleon
coupling constant and $f_\pi$ the pion weak decay constant. Note that
we assume this potential to be valid for any strictly positive
distance, $ r\neq 0$, so the limit $r \to 0^+ $ will be carefully
taken, {\it without} subtracting any contribution in the potential.

It is convenient to define the length scale
\begin{eqnarray}
R =   \frac{3 g_A^2 M }{32 \pi f_\pi^2}  
\label{eq:R_def} 
\end{eqnarray} 
which value is around 1 fm. For numerical calculations we take $f_\pi
=92.4 {\rm MeV}$, $M = 938.918 {\rm MeV}$, $ g_A =1.29 $ and hence $R
= 1.07764 {\rm fm}$. Using the Goldberger-Treiman relation, $g_{\pi
NN}= g_A M /f_\pi $, the corresponding pion nucleon coupling constant
is $ g_{\pi NN}=13.1083$ according to a phase shift analysis of NN
scattering~\cite{deSwart:1997ep}. Nevertheless, after the latest
determinations from the GMO sum rule~\cite{Ericson:2000md} we will
also take the value $ g_{\pi NN} =13.3158$. As we will see this
variation at the $5 \% $ level dominates the uncertainties in the OPE
calculations.

\subsection{The short distance regular solutions} 
\label{sec:short0} 

We look for normalized functions of the Eqs.~(\ref{eq:sch_coupled}),
\begin{eqnarray}
1= \int_0^\infty ( u(r)^2 + w(r)^2 ) dr  \, , 
\end{eqnarray} 
from which $A_S$ can be determined. The normalization at infinity is
guaranteed due to the asymptotic conditions,
Eq.~(\ref{eq:bcinfty_coupled}). However, the coupled channel potential
becomes singular at short distances, since $ U_T \to - 2R /r^3
$. Keeping only this term in Eqs.~(\ref{eq:sch_coupled}) one can
decouple the equations through the unitary
transformation~\cite{Martorell94}
\begin{eqnarray}
u_A (r) &=& \sqrt{\frac23} u (r)  + \frac{1}{\sqrt{3}}w (r) \, , \nonumber  \\
u_R (r) &=& -\frac1{\sqrt{3}}u (r) + \sqrt{\frac23} w (r) \, , 
\label{eq:eigenvectors}
\end{eqnarray} 
yielding an attractive singular potential $U_A \to -4 R/r^3 $ for
$u_A$ and $U_R \to 8 R/r^3 $ for $u_R$. Any solution obtained by
integrating from infinity with the Eq.~(\ref{eq:bcinfty_coupled}) down
to the origin has the asymptotic short distance
behavior~\footnote{The solutions for $u_A $ and $u_R$ are written in
terms of spherical Bessel functions~\cite{Martorell94}. We keep the
leading short distance behavior only.},
\begin{eqnarray}
u_R (r) &\to & \left(\frac{r}{R}\right)^{3/4} \left[ C_{1R} e^{+ 4
\sqrt{2} \sqrt{\frac{ R}{r}}} + C_{2R} e^{- 4 \sqrt{2} \sqrt{\frac{
R}{r}}} \right] \, , \nonumber \\ \\ u_A (r) &\to &
\left(\frac{r}{R}\right)^{3/4} \left[ C_{1A} e^{- 4 i \sqrt{\frac{
R}{r}}} + C_{2A} e^{ 4 i\sqrt{\frac{ R}{r}}} \right] \, . \nonumber
\label{eq:short_bc}
\end{eqnarray} 
The constants $C_{1R}$, $C_{2R}$, $C_{1A}$ and $C_{2A}$ depend
on both $\gamma $ and $\eta$ and the OPE potential parameters, $g_{\pi
NN} $ and $m$. Note that the leading short distance $r$ dependence
does not involve the pion mass and the deuteron wave number. Higher
order corrections to these solutions can be computed systematically to
high orders and are presented below in Sect.~\ref{sec:short}.

The regular solution at infinity contains the normalization
constant $A_S$, which is customarily set to one for computational
purposes, the deuteron wave number $ \gamma $ and the asymptotic D/S
ratio parameter $\eta$. The normalizability of the wave function at the
origin requires
\begin{eqnarray}
C_{1R} ( \gamma , \eta ) =0 \, ,  
\label{eq:c1r}
\end{eqnarray}  
which is a relation between $\eta$ and $\gamma$. The other remaining
constants are then completely fixed. This means that for the OPE
potential, the deuteron binding energy can be used as an independent
parameter. Thus, one has three independent variables, $ \gamma$,
the coupling constant with length scale dimension $R $ (or
equivalently $ g_{\pi NN }$) and the pion mass $ m$. Obviously, this
suggests integrating in from infinity and determining $\eta$ from the
regularity condition at the origin (\ref{eq:c1r}).

To analyze whether some additional condition arises let us check the
selfadjointness of the coupled channel Hamiltonian. The flux
at a point $r$ is given by
\begin{eqnarray}
i J (r) &=& u^* (r)' u(r) - u^* (r) u'(r) \nonumber \\
&+& w^* (r)' w(r) - w^* (r) w'(r)   \, , 
\end{eqnarray} 
so that current probability conservation at the origin implies
\begin{eqnarray}
| C_{1A}|^2-|C_{2A}|^2 =  2\sqrt{2} i \left(
C_{1R}^* C_{2R} - C_{2R}^* C_{1R} \right)  \, . 
\end{eqnarray} 
Thus, if we set $ C_{1R}=0$ there is no condition on $ C_{2R}$ and one
has $ C_{1A} = C_A e^{i \varphi} $ and $ C_{2A} = C_A e^{-i \varphi} $
with $C_A $ and $ \varphi$ real. So, we have three constants, $ C_{2R}
(\gamma) $, $C_A (\gamma) $ and $\varphi(\gamma)$, characterizing the
normalizable solutions at short distances for a given value of the
deuteron wave number $\gamma$, 
\begin{eqnarray}
u_R (r) &\to & C_{R} (\gamma) \left(\frac{r}{R}\right)^{3/4} e^{- 4
\sqrt{2} \sqrt{\frac{ R}{r}}}  \, ,  \nonumber \\ 
u_A (r) &\to & C_{A}(\gamma) \left(\frac{r}{R}\right)^{3/4} \sin \left[ 4
\sqrt{\frac{ R}{r}} + \varphi(\gamma) \right] \, . 
\label{eq:short_bc_reg}
\end{eqnarray} 
Actually, if we have any other state, say a scattering state with
positive energy, unitarity (i.e. orthogonality) requires that the
constant $\varphi(k) $ coincides with the bound state phase
$\varphi(\gamma)$. We will come back to this issue later when
discussing low energy parameters and scattering solutions in
Sect.~\ref{sec:scattering}. It is natural to expect that some
combination of short distance constants is independent on the OPE
potential parameters as they encode short distance physics. In
Sect.~\ref{sec:m_dependence} we establish, by demanding the standard
Feynman-Hellmann theorem, that specifically the short distance phase
$\varphi $ does not depend on the OPE potential parameters. In
Sect.~\ref{sec:short} we determine the values of the three constants
characterizing the three regular solutions by a detailed short
distance analysis of the OPE deuteron wave functions.

Note that any additional condition would actually {\it predict} both
$\gamma$ and $\eta$ from $m$ and $R$. This contradicts the claim of
Ref.~\cite{Martorell94} that $C_{2R} =0$, a conclusion implicitly used
in Ref~\cite{Beane:2001bc} and supported by the large $N_c$ argument
of Ref.~\cite{Beane:2002ab}. On the other hand, if one takes the
experimental values of $\eta$ and $\gamma$ as done in
Ref.~\cite{Phillips:1999am} one obtains both non vanishing $C_{1R}$
and $C_{2R}$ i.e., the irregular non-normalizable solution, unless a
short distance cut-off, $ R > 0.8 {\rm fm}$, is introduced as a physical
scale and not as an auxiliary removable regulator.

\subsection{Regularization with boundary conditions} 
\label{sec:cut-off}

Ideally, one would integrate in the large asymptotic solutions,
Eq.~(\ref {eq:bcinfty_coupled}), and match the short distance
behavior of Eq.~(\ref{eq:short_bc}) imposing the regularity condition
(\ref{eq:c1r}). In practice, however, the converging exponential at
the origin is rather elusive since integrated-in solutions quickly
run into the diverging exponentials due to round-off errors for $ r
\sim 0.05 {\rm fm}$ and dominate over the converging exponential. The
reason has to do with the fact that the natural scale where both
exponentials are comparable is rather large $r = 4 \sqrt{2} R \sim 6
{\rm fm}$, but in that region the lowest order short distance
approximation does not hold.

Instead, we will also try putting several short distance boundary
conditions corresponding to the choice of regular solutions at the
origin, 
\begin{eqnarray}
u(a)  &=& 0  \qquad ({\rm BC1}) \, , \nonumber \\  
u'(a) &=& 0  \qquad ({\rm BC2}) \, , \nonumber \\  
w(a) &=& 0  \qquad ({\rm BC3}) \, , \nonumber \\  
w'(a)  &=& 0  \qquad ({\rm BC4}) \, , \nonumber \\  
u(a) - \sqrt{2} w(a) &=& 0 \qquad ({\rm BC5}) \, , \nonumber  \\  
u'(a) - \sqrt{2} w'(a) &=& 0 \qquad ({\rm BC6}) \, , \nonumber \\  
\label{eq:bc_a}
\end{eqnarray}  
The advantage of using this kind of short distance cut-offs based on a
boundary condition is that there is only a single scale in the problem
as one naturally expects, and that one never needs to declare what is
the wave function below the boundary radius. Putting a square well
potential as a counter-term~\cite{Beane:2001bc} with depth $U_0 $
appears natural from standard perturbative experience but needs
specification of a further length scale, $1/\sqrt{U_0}$, and moreover,
generates multi-valuation
ambiguities~\cite{PavonValderrama:2003np,PavonValderrama:2004nb}.

\begin{figure}[]
\begin{center}
\epsfig{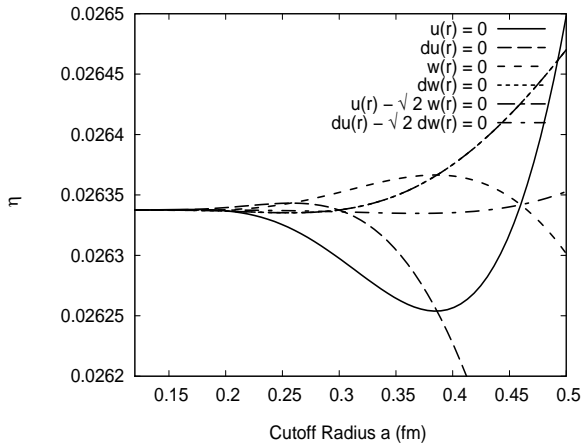}
\end{center}
\caption{The dependence of the asymptotic $D/S$ ratio $\eta $ on the
cut-off radius $a$ for several boundary conditions. We use $m=138.03 {\rm MeV} $ and $ R = 1.07764 {\rm fm}$ (corresponding to $ g_{\pi NN } = 13.1083 $).}
\label{fig:eta_a}
\end{figure}

It is convenient to use the superposition principle of boundary
conditions to write
\begin{eqnarray}
u (r) &=& u_S (r) + \eta \, u_D (r) \nonumber \\ w (r) &=& w_S (r) + \eta\,
w_D (r) \, , 
\end{eqnarray}
where $(u_S,w_S)$ and $(u_D,w_D)$ correspond to the boundary
conditions at infinity, Eq.~(\ref{eq:bcinfty_coupled}) with $A_S=1$
and $A_D=0$ and with $A_S=0$ and $A_D=1$ respectively. Thus, at the
boundary we can impose any of the conditions by just eliminating
$\eta$~\footnote{Numerically we find at the cut-off boundary $r=0.2 {\rm fm}$ 
\begin{eqnarray}
u(0.2) &=& 1139.23 - 43263.2\,\eta \nonumber \\
w(0.2) &=& -1807.33 + 68632.5\,\eta \nonumber \\
u'(0.2) &=& -35529.8 + 1.34913 \cdot \,{10}^6\,\eta \nonumber \\  
w'(0.2) &=& 55194.3 - 2.09606 \cdot \,{10}^6\,\eta  
\end{eqnarray}  
These large numbers appear because the of the dominance of the
diverging exponential at short distances.}. The resulting $\eta $
value obtained by all these boundary conditions is presented in
Fig.~\ref{fig:eta_a}. Actually, we see that the boundary condition $
u'(a) - \sqrt{2} w'(a) $ is about the smoothest condition we can think
of, since the $u_R $ combination goes to zero at small distances, its
derivative, $u_R' $ also goes to zero, although a bit less faster
since $u_R' / u_R \sim 1/r^{3/2}$. We see that all determinations of
$\eta$ based on any of the proposed cut-offs yield the same value with
great accuracy at cut-off radii below $0.2 {\rm fm} $. This is
somewhat fortunate since arithmetic precision is outraged typically
for $r< 0.06 {\rm fm}$. Obviously, any short distance cut-off
generates finite cut-off effects in the wave functions for distances
close to the cut-off radius. In Sect.~\ref{sec:short} we analyze this
problem by matching solutions of the form of Eq.~(\ref{eq:short_bc_reg})
to the integrated in numerical solutions and find that for many
practical purposes these finite cut-off effects are negligible. Thus,
we will base most of our results on the ``smoothest'' condition BC6 of
Eq.~(\ref{eq:bc_a}).

\medskip

\begin{figure}[]
\begin{center}
\epsfig{figure=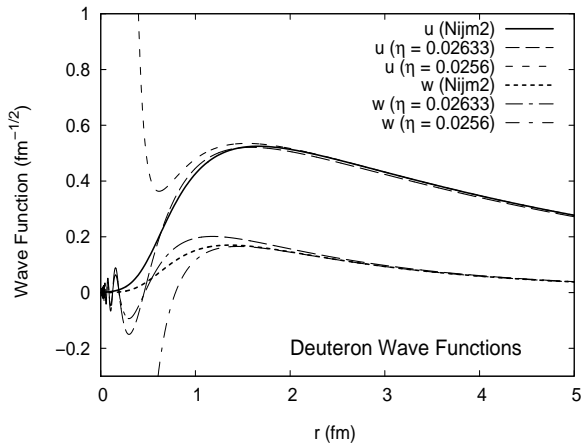,height=6cm,width=8cm}
\end{center}
\caption{The deuteron wave functions $u$ and $w$, obtained by
integrating in from infinity to the origin the OPE potential, compared
to those obtained with the Nijm II potential~\cite{Stoks:1994wp}. 
The irregular solutions
are obtained with the experimental $D/S$ ratio, $\eta_d = 0.0256 $, and
the regular ones with the OPE $D/S$ ratio, $\eta_{\rm OPE} =
0.026333$. We use $m=138.03 {\rm MeV} $ and $ R = 1.07764 {\rm fm}$
(corresponding to $ g_{\pi NN } = 13.1083 $).}
\label{fig:irregular+regular}
\end{figure}

The resulting deuteron wave functions $u$ and $w$ obtained by
integrating in from infinity to the origin the OPE potential are
plotted in Fig.~\ref{fig:irregular+regular} where the irregular
solutions are obtained with the experimental $D/S$ ratio $\eta_d =
0.0256 $ and the regular ones with the OPE $D/S$ ratio $\eta_{\rm
OPE} = 0.026333$. For comparison we also plot the NijmII deuteron wave
functions. We emphasize that the value of $\eta$ is a direct
consequence of taking the OPE down to the origin seriously.

\subsection{Deuteron observables} 

Once the solutions are known we can determine several observables of
interest. The matter radius reads,
\begin{eqnarray}
r_m^2 = \frac14 \langle r^2 \rangle = \frac14 \int_0^\infty r^2 ( u(r)^2 +
w(r)^2 ) dr 
\end{eqnarray} 
while potential contribution to the quadrupole moment (without meson
exchange currents) 
\begin{eqnarray}
Q_d  = \frac1{20} \int_0^\infty r^2 w(r) ( 2\sqrt{2} u(r)-w(r) ) dr  
\end{eqnarray} 
An important observable is the deuteron inverse radius 
\begin{eqnarray}
\langle r^{-1} \rangle = \int_0^\infty dr \frac{u(r)^2 + w(r)^2}{r}  
\end{eqnarray} 
which appears in low energy pion-deuteron scattering.  Finally, the
$D$-state probability is given by 
\begin{eqnarray}
P_D = \int_0^\infty w(r)^2  dr 
\end{eqnarray} 
Both $P_D $ and $\langle r^{-1} \rangle $ are sensitive to the
intermediate distance region around $2 {\rm fm}$ whereas $Q_d $ and
$r_m$ get their contribution from larger distances $\sim 4 {\rm fm} $.

The results for the asymptotic S-wave normalization $A_S$, the matter
radius $r_m$, the quadrupole moment, $Q_d$, and the D-state
probability, $P_D$ are presented in Table~\ref{tab:table_pert}. The
errors in the numerical calculation have been assessed by varying the
short distance cut-off in the range $a=0.1-0.2 {\rm fm}$ (in momentum
space that would naively correspond to take $ \Lambda =1/a = 2-4 {\rm
GeV}$). As we see, the cut-off uncertainty is smaller than the one
induced by variations at the $2\%$ level in the $g_{\pi NN}$ coupling
constant in the range between the lowest value ($\sim$ 13.1) obtained by a
fit to NN phase-shifts~\cite{deSwart:1997ep} and the highest recent
value ($\sim$ 13.3) determined from the GMO sum
rule~\cite{Ericson:2000md}. Equivalently, this uncertainty corresponds
to take $R=1.0776 {\rm fm} $ and $R = 1.1108 {\rm fm}$
respectively. Our results are generally speaking in agreement with
previous determinations where different sorts of cut-off methods have
also been implemented.

\subsection{Discussion}

At this point it may prove useful to ponder on the previous results
from a wider perspective. Let us remind that the basic assumption of an 
EFT is that the study of long wave length phenomena such as low energy
scattering or weakly bound systems do not require a detailed knowledge
of short distance physics. This general and widely accepted principle
requires some qualification because attractive and repulsive singular
potentials behave quite differently in this respect. Singular
attractive potentials, $\sim 1/r^n $, generate wave functions vanishing
as a power law, $r^{n/4} \sin ( r^{-n/2+1} + \varphi) $, and which need
a mixed boundary condition to specify the short distance phase
$\varphi$. Thus, short distance details become less important,
regardless on the value of $\varphi$. On the contrary, for singular
repulsive potentials the wave functions behave as 
$ r^{n/4} e^{\pm r^{-n/2+1}}$ and only
for the regular solution short distance details become irrelevant. In
the OPE potential, it is precisely the repulsive short distance OPE
component which requires a fine tuning of the solutions and eliminates
one {\it a priori} independent parameter like, e.g., the  asymptotic D/S
ratio $\eta$. As we see, if $\eta $ is treated as an independent
variable the short distance behavior of the deuteron wave functions
precludes the definition of a normalizable state due to the onset of
the {\it irregular} solution. This short distance insensitivity at low
energies could only be implemented by keeping the experimental $\eta$
value and {\it ignoring} OPE physics below some scale. The lower limit
established in Ref.~\cite{Phillips:1999am} to obtain a normalizable
state was $a\sim 1.3 {\rm fm}$ for the OPE potential. This obviously
requires some extension of the wave function below that scale and the
pretended model independence becomes a bit obscured. Our point is that
the short distance insensitivity materializes automatically for the
{\it regular} OPE deuteron wave functions since they vanish at the
origin.

\medskip

\begin{table*}
\caption{\label{tab:table_pert} Deuteron properties for the OPE
compared to the short range approximation and first order perturbation
theory. We use the non-relativistic relation $ \gamma= \sqrt{ 2
\mu_{np} B} $ with $B=2.224575(9)$ and take $m=138.03 {\rm MeV} $ and
$R= (3 /8 M ) g_{\pi NN}^2 / (4 \pi) = 1.07764 {\rm fm}$ corresponding
to $g_{\pi NN} =13.1083 $ \cite{deSwart:1997ep} (except the row OPE$^*$
where the value $g_{\pi NN}=13.316$~\cite{Ericson:2000md} has been
taken). The error is estimated by changing the short distance cut-off
in the range $a=0.1-0.2 {\rm fm} $.}
\begin{ruledtabular}
\begin{tabular}{|c|c|c|c|c|c|c|c|c|c|c|c|}
\hline & $\gamma ({\rm fm}^{-1})$ & $\eta$ & $A_S ( {\rm fm}^{-1/2}) $
& $r_m ({\rm fm})$ & $Q_d ( {\rm fm}^2) $ & $P_D $ & $\langle r^{-1}
\rangle $ & $ \alpha_0 ({\rm fm}) $ & $\alpha_{02} ({\rm fm}^3) $ & $
\alpha_2 ({\rm fm}^5) $ & $r_0 ({\rm fm} ) $ \\ \hline
{\rm Short} & Input & 0 & 0.6806 & 1.5265 & 0 & 0\% & $\infty $ & 4.3177& 0 & 0 & 0   \\ 
{\rm OPE}(pert) & Input & 0.051 & 0.7373 &  1.6429 & 0.4555
& 0\% & $\infty $ & 4.6089  &  2.5365  & 0  &  0.4831 \\ 
{\rm OPE} & Input & 0.02633 & 0.8681(1) & 1.9351(5) & 0.2762(1)
& 7.88(1)\% & 0.476(3) & 5.335(1) & 1.673(1) & 6.169(1) & 1.638(1) \\ 
{\rm OPE}$^*$ & Input & 0.02687 & 0.8718(2) & 1.9429(6) & 0.2826(2)
& 7.42(1)\% & 0.471(3) & 5.353(1) & 1.715(1) & 6.4001(1) & 1.663(1) \\ 
\hline 
NijmII & Input & 0.02521 & 0.8845(8) & 1.9675 & 0.2707 & 
5.635\% &  0.4502 & 5.418 & 1.647 & 6.505 & 1.753 \\
Reid93 & Input & 0.02514 & 0.8845(8) & 1.9686 & 0.2703 & 
5.699\% & 0.4515 & 5.422 & 1.645 & 6.453 & 1.755 \\ \hline 
Exp. \footnotemark[1] &  0.231605 &  0.0256(4)  & 0.8846(9) & 1.9754(9)  &
0.2859(3) & 5.67(4)  & & 5.419(7) & & & 1.753(8) \\ \hline 
\end{tabular}
\end{ruledtabular}
\footnotetext[1]{(Non relativistic). See
e.g. Ref.~\cite{deSwart:1995ui} and references therein.}
\end{table*}

\section{OPE Correlations in deuteron observables} 
\label{sec:OPE-corr} 

As we have said, in the OPE potential we can use the deuteron wave
number as an input of the calculation on the same footing as $g_{\pi
NN} $ and the pion mass $m$. Then, other observables are
predicted. We will study now the dependence of these observables on
$\gamma$, $m $ and $R$. 

\subsection{Dependence on the Binding Energy}

\medskip
\begin{figure*}[]
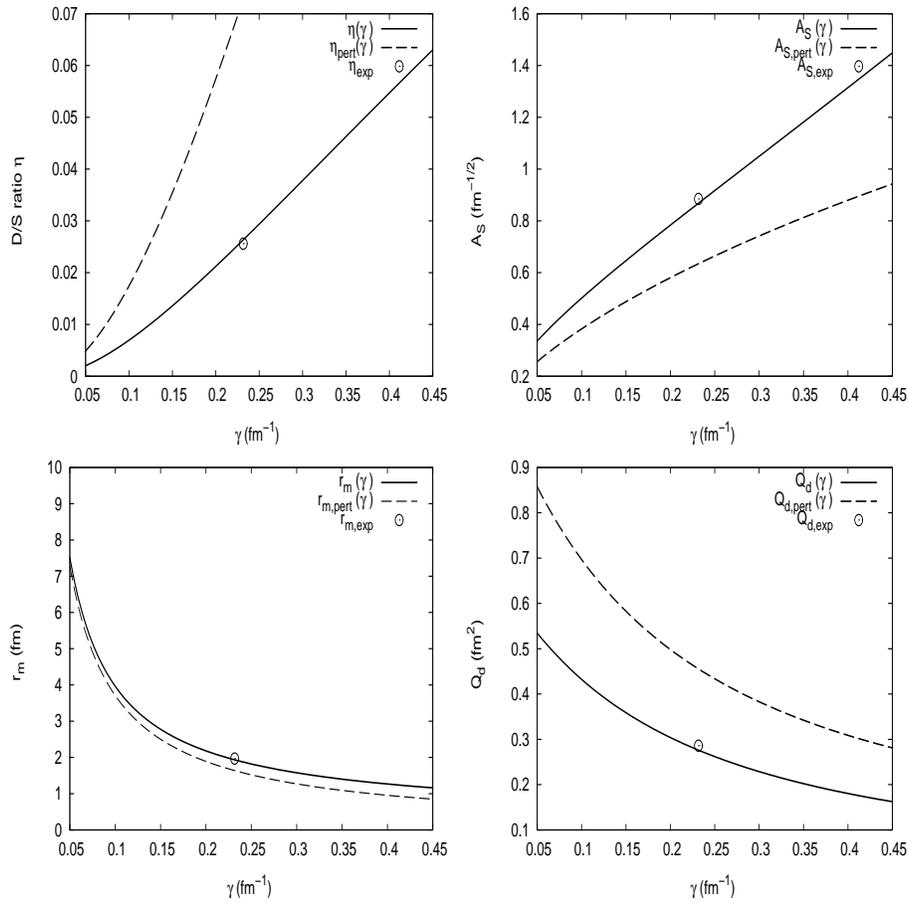

\begin{center}
\epsfig{figure=eta-gamma.epsi,height=6cm,width=6cm}
\epsfig{figure=As-gamma.epsi,height=6cm,width=6cm}\\ 
\epsfig{figure=rm-gamma.epsi,height=6cm,width=6cm}
\epsfig{figure=Qd-gamma.epsi,height=6cm,width=6cm}
\end{center}
\caption{The dependence of the asymptotic $D/S$ ratio $\eta $ (upper
left panel), the S-wave normalization $A_S$ (in ${\rm fm}^{-1/2} $,
upper right panel) , the matter radius $r_m$ (in ${\rm fm} $, lower
left panel) and the quadrupole moment $ Q_d $ (in ${\rm fm}^2 $,
lower right panel) on the deuteron wave number $\gamma$ (in ${\rm
fm}^{-1}$) for the short range theory, the first order perturbative
result and the exact OPE result.  The points represent the
experimental values. We use $m=138.03 {\rm MeV}$ and $ R=1.07764 {\rm fm}
$.}
\label{fig:eta[gamma]}
\end{figure*}

In Fig.~\ref{fig:eta[gamma]} we show the dependence of the D/S ratio
as a function of the deuteron wave number $\gamma$ keeping $m$ and $R$
fixed. In the weak binding limit $\gamma \ll m_\pi $, long distances
dominate and the finiteness of the wave function at a point $ r \gg
1/m $ requires $\eta \sim \gamma^2 $. The radius of convergence of
such an expansion for the observables is $|\gamma| < m /2 $, since the
integrals involve the factor $e^{-(2 \gamma + m) r}$ at large
distances, diverging for $ \gamma < -m/2 $. The experimental number
is not far from $ \gamma = m/3 $, which is within the domain of
analyticity but somewhat close to the convergence radius. So, one
expects a slow convergence. As we see in the weak binding limit we
have a quadratic behavior $ \eta_{\rm OPE} \sim \gamma^2 $ whereas
for stronger binding a linear behavior sets in. It is remarkable that
the experimental values in the intermediate regime.  On the other
hand, in the strong binding case $\gamma \gg m_\pi $, short distances
dominate and we must have $ \eta \sim 1/\sqrt{2}$.  Numerically we
find for the deuteron observables, 
\begin{eqnarray}
\eta^{\rm OPE} &=& 0.9638 \gamma^2 - 3.46864 \gamma^3+ {\cal O}
(\gamma^4) \\ \frac{A_S^{\rm OPE}}{\sqrt{2 \gamma}} &=& 1 + 1.2455
\gamma -0.4705 \gamma^2 + {\cal O}(\gamma^3) \\ 
\sqrt{8} \, \gamma \,\, r_m^{\rm OPE} 
&=& 1 + 1.2455 \gamma -0.4705 \gamma^2 + {\cal O}
(\gamma^3) \\ 
Q_d^{\rm OPE} &=& 0.6815 - 3.5437 \gamma + {\cal O} (\gamma^2)
\end{eqnarray} 
Note that we have the weak binding correlation
\begin{eqnarray}
r_m = \frac{A_S}{4\,\gamma^{3/2}} + {\cal O} ( \gamma^3 )    
\end{eqnarray} 
which is compatible at the $2 \sigma $ confidence level with data; for
the experimental value $A_S= 0.8845 (8) $ the value $r_m=1.984(2) $ to
be compared with the experimental number $r_m=1.971(6) $. In the weak
binding limit one also has the correlation
\begin{eqnarray}
\frac{\sqrt{2} \gamma^2 Q_d}{\eta_d}  = 1 + {\cal O} (\gamma)  
\end{eqnarray}
a dependence that one would expect on general grounds by just taking
the asymptotic formulas and neglecting the $w(r)^2$ term in the
expression for the quadrupole moment. Experimentally this relation is
fulfilled with a $15\%$ accuracy. The non-perturbative OPE value is
actually closer to potential models. 

\subsection{Comparison with perturbation theory} 

It is instructive to solve the coupled deuteron equations,
Eq.~(\ref{eq:sch_coupled}) in standard perturbation theory for the
fixed energy bound state. One of the reasons is to check the
correctness of our non-perturbative calculations in the weak binding
regime. Another motivation is to establish contact with the
perturbative calculations of Ref.~\cite{Kaplan:1998sz} where
dimensional regularization in the power divergence subtraction (PDS)
scheme was implemented. Finally, there is the question of
quantitatively assessing the validity of such an approximation. We
relegate the calculation to Appendix~\ref{sec:pert}. At first order in
perturbation theory one gets in the weak binding limit
\begin{eqnarray}
  \eta_{\rm pert} &=& 1.5497 \gamma^2 -4.15479 \gamma^3 + {\cal O} ( \gamma^4, R^2)  \nonumber \\ 
  \frac{A_{S,{\rm pert}}}{\sqrt{2 \gamma}} &=& 1 - 0.7184 \gamma -
    2.7394 \gamma^2 + {\cal O} ( \gamma^3, R^2)\nonumber \\
r_{m,\rm pert}  \sqrt{8} \gamma &=& 1 + 0.71843 \gamma - 2.7394
\gamma^2 + {\cal O} ( \gamma^3, R^2) \nonumber \\
Q_{\rm pert} &=& 1.09587 - 5.87576 \gamma + {\cal O} ( \gamma^2, R^2)
\nonumber \\
\end{eqnarray} 
The nominally ${\cal O} (R^2)$ second order contributions are in fact
divergent because the leading order correction to the D-wave component
$w(r) $ diverges at the origin (see Appendix~\ref{sec:pert}). In
general terms we find that the exact OPE results are estimated within
$ 30\%$ by the first order perturbative calculations of
Appendix.~\ref{sec:pert}.

\section{dependence on the pion mass and chiral limit} 
\label{sec:m_dependence}

Recent works~\cite{Bulgac:1997ji,
Beane:2002xf,Beane:2002vs,Epelbaum:2002gb} predict the change of the
deuteron binding energy as a function of the pion mass by taking the
experimental binding energy at the physical value of the pion mass and
making the additional assumption short distance physics to be
independent on the pion mass. While it is true that the leading short
distance $r$ dependence of the deuteron wave functions are independent
on the pion mass, the constants $C_A$, $ C_R $ and $ \varphi $ do in
principle depend on the three independent parameters $ m$, $g_{\pi NN}
$ and $\gamma$.  As we have noted $\gamma $ cannot be predicted for
the OPE potential.  So the approach pursued in
Refs.~\cite{Bulgac:1997ji, Beane:2002xf,Beane:2002vs,Epelbaum:2002gb}
is equivalent to integrate in with the physical pion mass and then
integrate out fixing some combination of short distance constants with
the unphysical pion mass and searching for the appropriate regular
solution at infinity. If one makes the pion lighter long distance
effects should dominate, and one could just use the OPE potential to
estimate the chiral limit as a first approximation. It is thus
interesting to analyze the pion mass dependence both explicitly
(i.e. varying $m$ in the OPE potential) and implicitly (i.e. taking
into account the dependence of the OPE coupling $R$ on the pions
mass). We will determine the pertinent combination of short distance
constants by demanding the Feynman-Hellmann theorem in the OPE
potential.

\subsection{Explicit pion mass dependence} 

To proceed, let us assume that to an infinitesimal change in $m \to m
+ \Delta m $ there corresponds a change both in the deuteron wave
number $\gamma \to \gamma + \Delta \gamma $ and in the coupled channel
potential matrix $ U (r) \to U(r)+ \Delta U(r) $. We can write a
Lagrange identity by varying the equation and its adjoint, yielding
for a normalized state
\begin{eqnarray}
- \frac{\partial \gamma^2}{\partial m} &=& \langle \Psi_m |
  \frac{\partial U }{\partial m} | \Psi_m \rangle \nonumber \\ &+&
  \left[ u' \frac{\partial u}{\partial m} - u \frac{\partial
  u'}{\partial m } + w' \frac{\partial w}{\partial m} - w \frac{\partial
  w'}{\partial m } \right] \Bigg|_0^\infty
\label{eq:chiral}
\end{eqnarray} 
This is an extended Feynman-Hellmann theorem where the second term in
the l.h.s. corresponds to the short distance contribution (the term at
infinity vanishes for a bound state). One of the advantages of the
Feynman-Hellmann theorem is that one could in principle establish
comparison theorems, provided the change in the coupled channel
potential matrix, $\Delta U$, is a definite quadratic form. Note also that
the derivative with respect to $m$ annihilates the centrifugal term,
$6/r^2 $, and one can diagonalize the coupled channel potential by the
unitary transformation Eq.~(\ref{eq:eigenvectors}) so that the result
behaves additively in the attractive and repulsive
eigenchannels. Using the leading short distance behavior,
Eq.~(\ref{eq:short_bc_reg}), we therefore get
\begin{eqnarray}
- \frac{\partial \gamma^2}{\partial m} &=& \int_0^\infty dr \left[
  u_A(r)^2 \frac{\partial U_A}{\partial m} + u_R(r)^2 \frac{\partial
  U_R}{\partial m} \right] \nonumber \\ &+& C_A^2 \frac{d \varphi}{d
  m} \, .
\end{eqnarray} 
As we see, assuming as suggested in Ref.~\cite{Bulgac:1997ji} that the
short distance physics does not depend on the pion mass corresponds to
demanding the standard Feynman-Hellmann theorem where only the OPE
potential change contributes. For $C_A \neq 0 $ one obtains the
condition
\begin{eqnarray}
\frac{d }{dm} \varphi (\gamma , m) = \frac{\partial \varphi}{\partial
\gamma} \frac{d \gamma}{d m} + \frac{\partial \varphi}{\partial
m} =0 \, ,   
\label{eq:varphi_m}  
\end{eqnarray} 
whence a functional relation between the pion mass and the deuteron
binding energy follows. However, note that even in this case the sign
of the result is indefinite since
\begin{eqnarray}
\frac{\partial \gamma^2}{\partial m} &=& \int_0^\infty dr \left[
  u_A(r)^2 \frac{\partial U_A}{\partial m} + u_R(r)^2 \frac{\partial
  U_R}{\partial m} \right] \, . 
\end{eqnarray} 
So, we have to determine the sign numerically. 

The relation (\ref{eq:varphi_m}) has an equivalent formulation in the
boundary condition regularization. For instance, if we assume the same
condition BC6 of Eq.~(\ref{eq:bc6}) for all values of the pion mass we
get
\begin{eqnarray}
\frac{d}{dm}\left[ \frac{w' (a) }{u (a)\sqrt{2} + w (a)} \right] =0 \, , 
\end{eqnarray} 
where $ a$ is taken to be independent of $m$. In practice, one
computes the ratio within the bracket for the physical pion mass and
searches for $\gamma$ such that the ratio for the unphysical pion mass
yields the same numerical value. If we take the chiral limit we get $
B_d ( 0 , g_{\pi NN} ) = 4.3539 {\rm MeV} $ This value is very close
to the one found in Ref.~\cite{Beane:2001bc} $ B_d ( 0 , g_{\pi NN} )
= 4.2 {\rm MeV}$~\footnote{These authors look for poles of the
$S-$matrix  so constructed as to reproduce the physical scattering
length $\alpha_0 = 5.42 {\rm fm} $ and effective range $r_0 = 1.75
{\rm fm}$ at the physical value of the pion mass in the $^3S_1$ eigen
channel. By doing so the explicit dependence on $g_{\pi NN}$ becomes
rather weak. Actually, they take $ g_{\pi NN}=12.73 $ and we would get
instead $ B_d ( 0 , g_{\pi NN} ) = 0.98 {\rm MeV} $ instead. This
apparent contradiction is resolved by noting that, as we will see
below, for $g_{\pi NN} = 13.1083$ in the OPE we get an scattering length
$\alpha_0 = 5.335 {\rm fm} $ an effective range of $r_0=1.63$ quite
close to the experimental values.}. Deuteron observables in the
explicit $ m=0 $ limit are listed in Table~\ref{tab:table0}.

\subsection{Implicit pion mass dependence} 

To take into account the implicit pion mass dependence we have to take
into account the dependence of $ R= 3g_A^2 M / 32 \pi f_\pi^2 $ on the
pion mass. In the chiral limit one gets a larger OPE
coupling~\cite{Epelbaum:2002gb}. The value is uncertain and as an
educated guess we take $R_0 =1.06(2) R $. Using the same formulation
as in the $m$ dependence, the change in the deuteron binding with
respect to the $g_{\pi NN} $ coupling constant or equivalently
the scale dimension $R$ we get (assuming as before the short
distance angle $\varphi $ to be independent on m),
\begin{eqnarray}
- R \frac{\partial \gamma^2}{\partial R } &=& \int_0^\infty dr \left[
  u_A(r)^2 U_A + u_R(r)^2 U_R \right]  \nonumber \\ 
\label{eq:dgamma/dR}
\end{eqnarray} 
Again, the result is indefinite since $U_A < 0 $ and $U_R >0 $ and it
is not obvious, unlike naive expectations, that a stronger coupling
provides stronger binding. The sign depends actually on the details of
the wave functions and the particular values of the parameters.
Numerically one finds $d \gamma / dR > 0 $, a trend that can be
understood if the repulsive term in Eq.~(\ref{eq:dgamma/dR}) is
neglected or on the basis of the inequality $|u_A| > |u_R| $ which is
numerically fulfilled. In any case one has the differential inequality
$ d \gamma / dR < \gamma / R $. 

Numerically, we get $ \gamma_0 = 0.61(10) {\rm fm}^{-1}$ and hence
\begin{eqnarray}
B_d^0 = 15 (5) {\rm MeV}  
\end{eqnarray} 
a value compatible with the analysis of Ref.~\cite{Epelbaum:2002gb}
$B_d^0 =9.6 \pm 3 $ (perhaps with larger
errors~\cite{Beane:2002xf}~\footnote{If we take $R_0 = 1.1 R $ we get
$B_d^0= 33 {\rm MeV}$.}) but in disagreement with
Refs.~\cite{Beane:2001bc,Beane:2002vs} where the deuteron becomes
unbound for $m < 90 {\rm MeV}$. In any case we confirm the trend of
having a stronger binding of the deuteron in the chiral limit. The
corresponding observables can be looked up in Table.~\ref{tab:table0}.

\medskip
\begin{figure}[]
\begin{center}
\epsfig{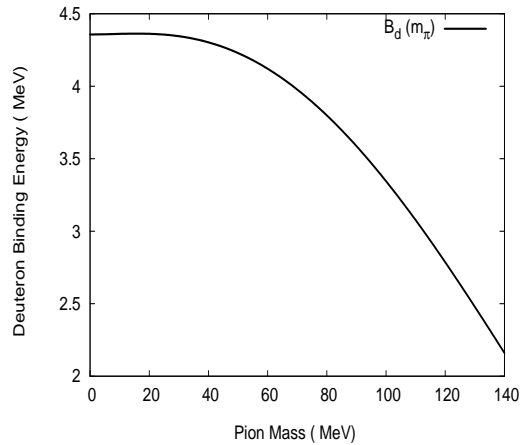}
\end{center}
\caption{Explicit pion mass dependence of the deuteron binding energy,
keeping $g_{\pi NN} = 13.1083 $ fixed. We have $ B_d(m) = 2.22457 {\rm
MeV} $, $ B_d (0) = 4.3539 {\rm MeV}$.}
\label{fig:Bd[m]}
\end{figure}

\begin{table*}
\caption{\label{tab:table0} Deuteron properties for the OPE and their
dependence on the pion mass.  We use the non-relativistic relation $
\gamma= \sqrt{ 2 \mu_{np} B} = 0.231605 \,{\rm fm}^{-1} $ with
$B=2.224575(9)$ for $m=138.03 $ and $g_{\pi NN}=13.1083$. $m=0$
(explicit) means taking $m=0$ but using $R = 1.07764 {\rm fm}$ (or
equivalently $g_{\pi NN}=13.1083$). $m=0$ (implicit) means taking $m=0$
and using $R_0 = 1.06(2) \times R $. }
\begin{ruledtabular}
\begin{tabular}{|c|c|c|c|c|c|c|}
\hline & $\gamma ({\rm fm}^{-1})$ & $\eta$ & $A_S ( {\rm fm}^{-1/2}) $
& $r_d ({\rm fm})$ & $Q_d ( {\rm fm}^2) $ & $P_D $ \\ \hline 
$m=138.03$ MeV & Input & 0.02633 & 0.8681(1) & 1.9351(5) & 0.2762(1)
& 7.88(1)\%  \\
$m=0$  (explicit) & 0.3240(1)  & 0.09452  & 0.8444(1) & 1.550(1) & 0.3006(3) 
& 10.96(2) \% \\
$m=0$  (implicit) & 0.61(10)  & 0.15(2) & 0.48(7) & 0.98(10) & 0.15(3)
& 15(1) \% \\ \hline  
\end{tabular}
\end{ruledtabular}
\end{table*}

\section{Scattering properties in the $^3S_1-{}^3D_1$ channel} 
\label{sec:scattering}  

\subsection{Orthogonality constraints and Phase Shifts} 
\label{sec:phase-shifts}

For the $\alpha$ and $\beta$ positive energy scattering states we
choose the asymptotic normalization
\begin{eqnarray}
u_{k,\alpha} (r) &\to & \frac{\cos \epsilon}{\sin \delta_1}\Big( \hat
j_0 (kr) \cos \delta_1 - \hat y_0 (kr) \sin \delta_1 \Big) \, , \nonumber \\ w_{k,\alpha}
(r) &\to & \frac{\sin \epsilon}{\sin \delta_1}\Big( \hat j_2 (kr) -
\hat y_2(kr) \sin \delta_1 \Big) \, , \nonumber \\ \\  
u_{k,\beta} (r) & \to & -\frac1{\sin \delta_1}\Big( \hat j_0 (kr) \cos \delta_2 - y_0 (kr)
\sin \delta_2 \Big) \, ,  \nonumber \\ 
w_{k,\beta} (r) &\to & \frac{\tan \epsilon}{\sin \delta_1}\Big( \hat
j_2 (kr) \cos \delta_2 - \hat y_2(kr) \sin \delta_2 \Big) \, , \nonumber \\ 
\end{eqnarray} 
where $ \hat j_l (x) = x j_l (x) $ and $ \hat y_l (x) = x y_l (x) $
are the reduced spherical Bessel functions and $\delta_1$ and
$\delta_2$ are the eigen-phases in the $^3S_1$ and $^3D_1$ channels, and
$\epsilon$ is the mixing angle $E_1$. Again, the general solution at
short distances is given by the general Eq.~(\ref{eq:short_bc_reg}),
where the constants $C_A $, $ C_R $ and $\varphi$ are now different
since we have a zero energy state and depend whether we have an
$\alpha$ or $\beta$ state, so we have the short distance constants, $
C_{A,\alpha} (k) , C_{R,\alpha} (k) , \varphi_\alpha (k) $ and $
C_{A,\beta} (k) , C_{R,\beta} (k) , \varphi_\beta (k) $ respectively.
This implies certain correlations between $\delta_1$, $\delta_2$ and
$\epsilon$.

For a regular self-adjoint potential the orthogonality of bound and
scattering states comes out automatically. We look now for the
consequences of {\it demanding} this property in the singular OPE
potential. Using the standard manipulations to prove orthogonality
between states of different energy we get the following relation
between $\alpha$ and $\beta$ states and the bound deuteron state
(which we denote by a subscript $\gamma$ in this section),
\begin{eqnarray}
0 &=& (\gamma^2+k^2 ) \int_0^\infty dr \Big[ u_\gamma (r) u_k (r) +
w_\gamma (r) w_k (r) \Big] \nonumber \\ &=& \left[ u_\gamma' u_k -
u_\gamma u_k' + w_\gamma' w_k - w_\gamma w_k' \right] \Big|_0^\infty
\label{eq:orth}
\end{eqnarray} 
Using the short distance solution, Eq.~(\ref{eq:short_bc_reg}), we get  
\begin{eqnarray}
C_{A,i} (k) C_A (\gamma) \sin\left[ \varphi (\gamma) - \varphi_i (k)
\right] =0 \qquad \, , \, i = \alpha ,\beta 
\end{eqnarray} 
Which yields 
\begin{eqnarray}
\varphi (\gamma) = \varphi_\alpha (k) = \varphi_\beta (k)
\label{eq:phi_const}
\end{eqnarray} 
Thus, the short distance phases $ \varphi_\alpha(k) $ and $
\varphi_\beta(k) $ of the $^3S_1-{}^3D_1$ channel wave functions in the
OPE potential at short distances are all determined by deuteron
properties. This means, in particular that the low energy parameters
and scattering phase shifts are uniquely determined by the deuteron
binding energy and the OPE potential parameters~\footnote{This property
does not hold for other triplet channels with higher partial waves,
because there are no bound states in those channels. Nevertheless, it
is also true that there is only one independent parameter. This means
in practice that one can use one scattering length out of the three to
predict the phase shifts also in other partial waves.}.

The previous argument can also be implemented if we have a short
distance cut-off at $r=a$, the orthogonality relation of
Eq.~(\ref{eq:orth}) transforms into the condition
\begin{eqnarray}
&& u_\gamma' (a) u_{k,i} (a) + w_\gamma' (a) w_{k,i} (a) \\ 
&=& u_\gamma (a)
u_{k,i}'(a) + w_\gamma (a) w_{k,i} '(a) \nonumber \\ && \quad
i=\alpha,\beta
\label{eq:orth_a}
\end{eqnarray} 
Thus, if we impose the same condition on both solutions,
Eq.~(\ref{eq:orth_a}) cannot be satisfied unless they are related at
the boundary. For instance for the condition BC6 of
Eqs.~(\ref{eq:bc_a}), we get the two relations 
\begin{eqnarray}
u_{k,i}' (a) = \sqrt{2} w_{k,i}' (a) \, , \qquad i=\alpha,\beta 
 \label{eq:bc6} 
\end{eqnarray} 
The orthogonality relation corresponding to boundary conditions of the form
of Eq.~(\ref{eq:bc6}) implies then the orthogonality constraint 
\begin{eqnarray}
\frac{w_{k,i}'(a)}{u_{k,i} (a)\sqrt{2} + w_{k,i} (a)} = \frac{w_\gamma' (a)
}{u_\gamma (a)\sqrt{2} + w_\gamma (a)}
\end{eqnarray} 
Which is the analog finite cut-off condition of
Eq.~(\ref{eq:phi_const}). The remaining conditions in
Eqs.~(\ref{eq:bc_a}) generate analogous orthogonality constraints.

\begin{figure*}[]
\begin{center}
\epsfig{figure=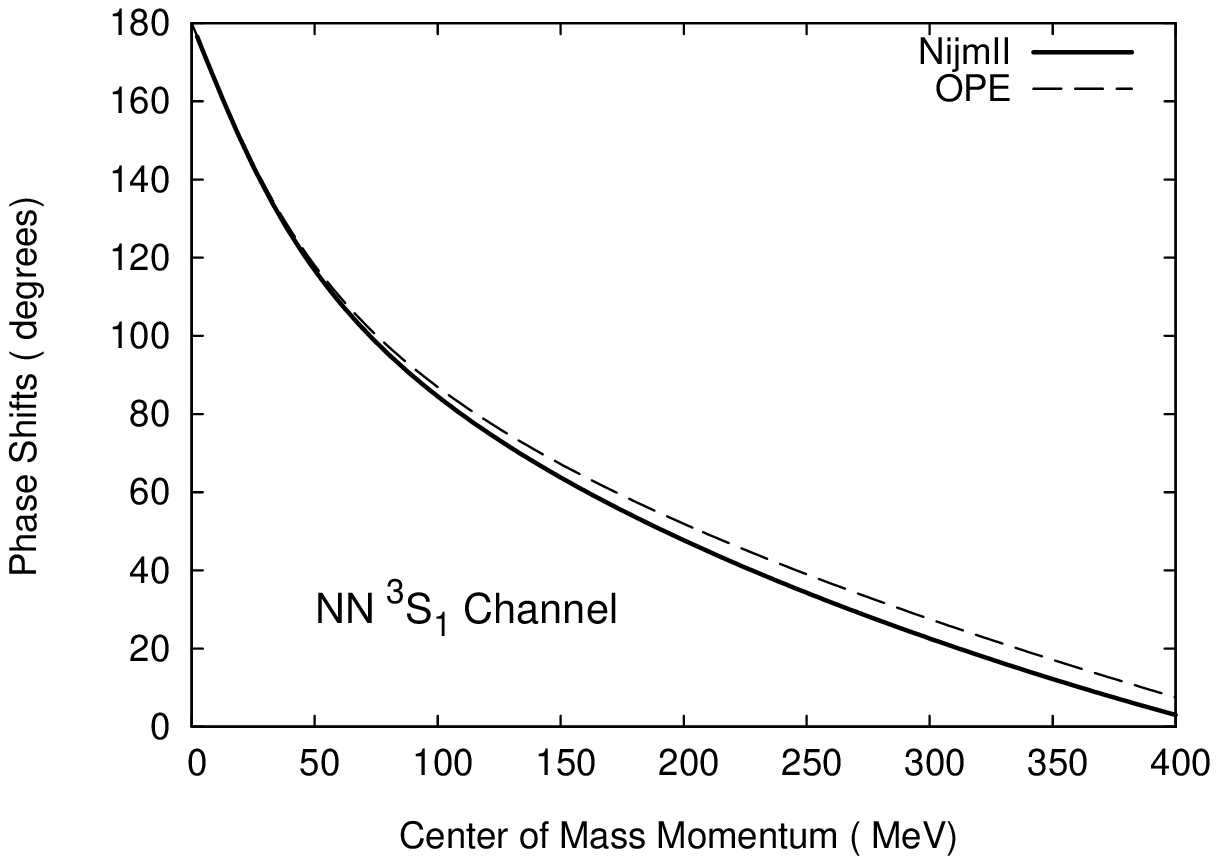,height=5.5cm,width=5.5cm}
\epsfig{figure=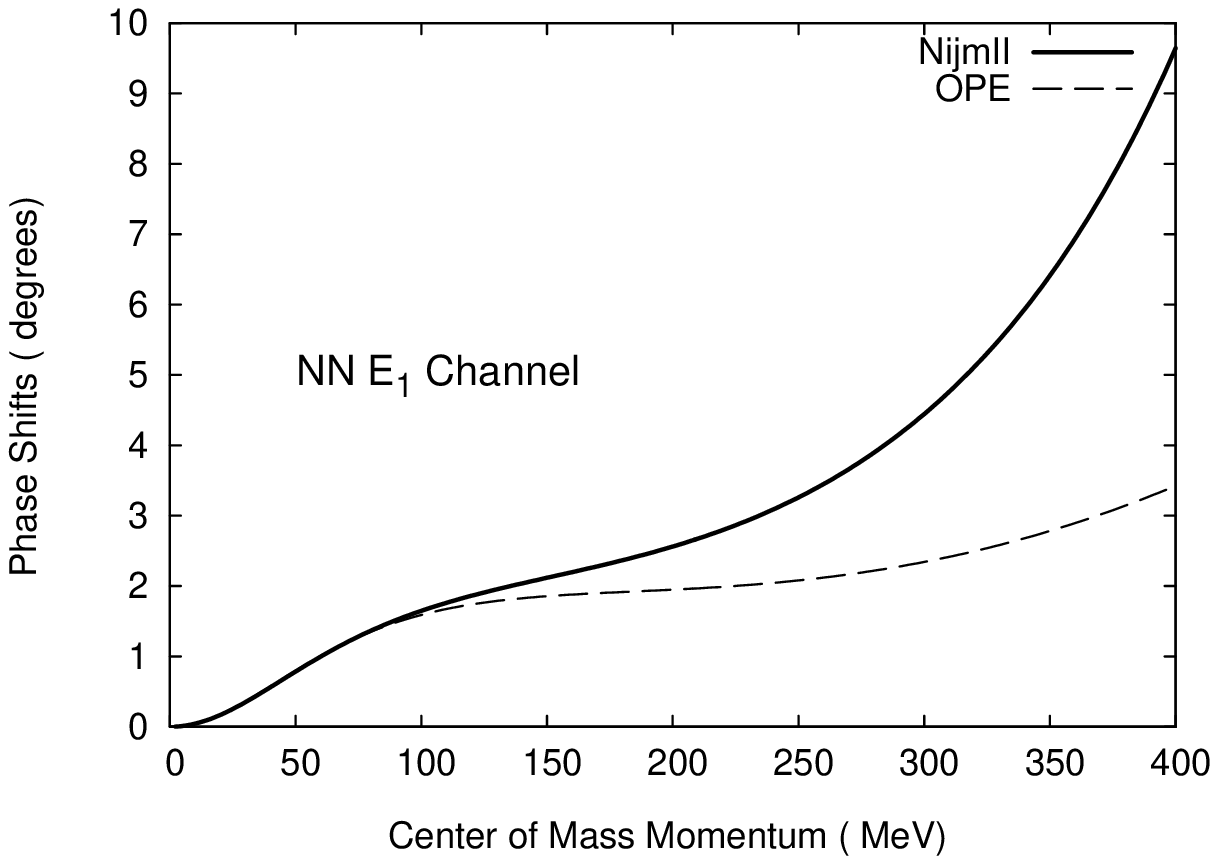,height=5.5cm,width=5.5cm}
\epsfig{figure=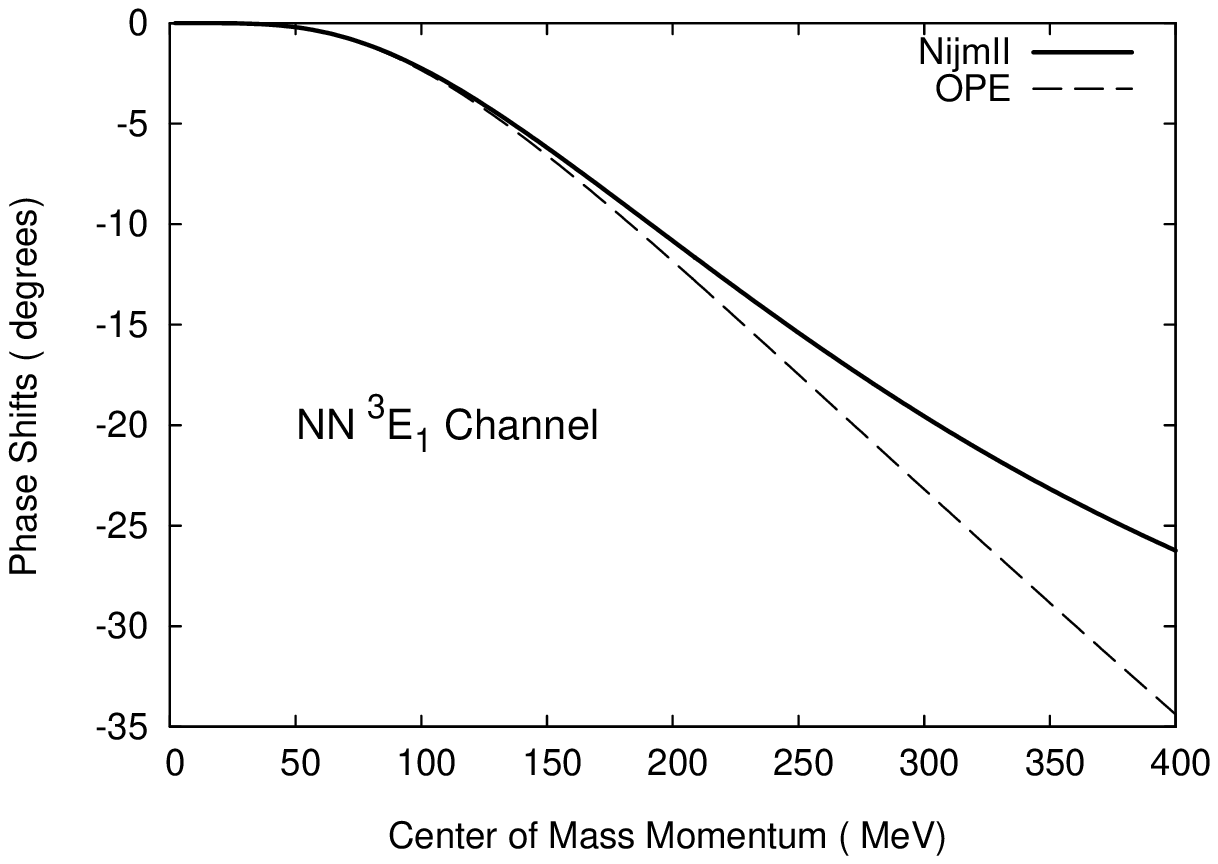,height=5.5cm,width=5.5cm}
\end{center}
\caption{Eigen Phase shifts for the OPE potential as a function of the
CM np momentum in the triplet $^3S_1-{}^3D_1$ channel compared to the
Nijmegen results~\cite{Stoks:1993tb}. The regular scattering wave
functions are orthogonal to the regular deuteron bound state wave
functions constructed from the OPE with $\gamma=0.231605 {\rm
fm}^{-1}$, $ m=138.03 {\rm MeV}$ and $ g_{\pi NN} = 13.1083 $.}
\label{fig:phase-shifts}
\end{figure*}

The results for the $^3S_1-{}^3D_1$ channel phase shifts using these
conditions are presented in Fig.~\ref{fig:phase-shifts}. The
description is rather satisfactory and it seems to work, as one might
expect, up to the vicinity of the CM momentum which magnitude
coincides with the two pion exchange left cut $ k={\rm i} \,m $.

\medskip

\subsection{Low energy parameters} 
\label{sec:low-energy} 

\begin{figure*}[]
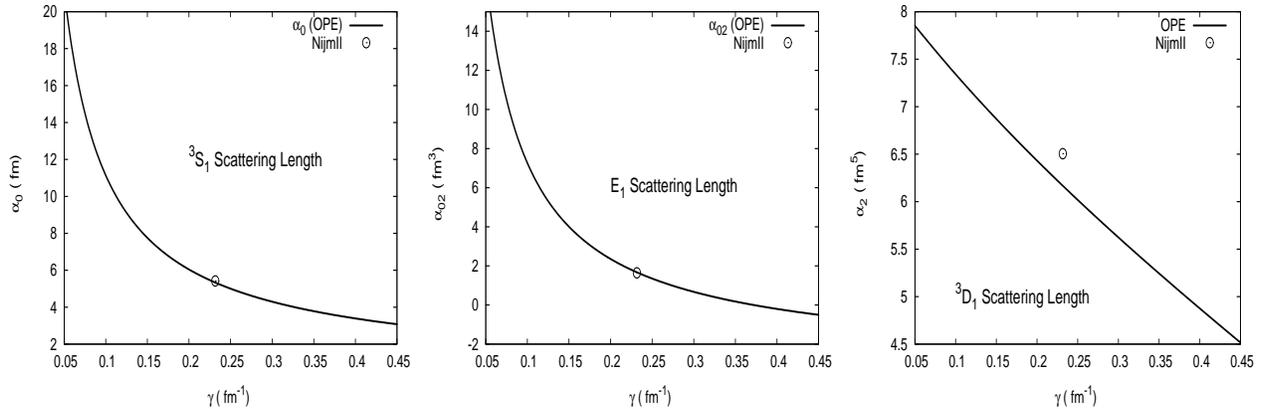

\begin{center}
\epsfig{figure=a0-gamma.epsi,height=5.5cm,width=5.5cm}
\epsfig{figure=a02-gamma.epsi,height=5.5cm,width=5.5cm}
\epsfig{figure=a2-gamma.epsi,height=5.5cm,width=5.5cm}
\end{center}
\caption{The dependence of the $^3S_1-{}^3D_1$ scattering lengths
$\alpha_0$ , $\alpha_{02}$ and $\alpha_2$ on the deuteron wave number
$\gamma$ (in ${\rm fm}^{-1}$). The point represents the experimental
values.}
\label{fig:a0[Gamma]}
\end{figure*}

In the low energy limit one has 
\begin{eqnarray}
\delta_1 &\to& - \alpha_0 k \, , \nonumber \\ 
 \delta_2 &\to & -\alpha_2 k^5  \, , \nonumber \\
\epsilon & \to & \frac{\alpha_{02}}{\alpha_0} k^2  
\end{eqnarray} 
so that the  zero energy the wave functions behave asymptotically 
\begin{eqnarray}
u_{0,\alpha} (r) & \to & 1- \frac{r}{\alpha_0} \, , \nonumber \\ w_{0,\alpha} (r) &
\to & \frac{3 \alpha_{02}}{\alpha_0 r^2 } \, , \nonumber \\ u_{0,\beta} (r)
&\to & \frac{r}{\alpha_0} \, , \nonumber \\ w_{0,\beta} (r) &=& \frac{3
\alpha_{2}}{\alpha_{02} r^2 }- \frac{r^3}{15 \alpha_{02}} \, . 
\label{eq:zero_energy}
\end{eqnarray} 
Using these zero energy solutions one can determine the effective
range. The $^3S_1$ effective range parameter is given by
\begin{eqnarray} 
r_0 &=& 2 \int_0^\infty \left[ \left(1-\frac{r}{\alpha_0} \right)^2 -
u_\alpha (r)^2 - w_\alpha (r)^2 \right] dr \, . \nonumber \\ 
\end{eqnarray} 
In the zero energy case, the vanishing of the diverging exponentials
at the origin imposes a condition on the $\alpha $ and $\beta$ states
which generate a correlation between $\alpha_0$ , $\alpha_{02}$ and
$\alpha_2$. Using the superposition principle of boundary conditions
we may write the solutions in such a way that 
\begin{eqnarray} 
u_{0,\alpha} (r) &=& u_1 (r) - \frac{1}{\alpha_0} u_2 (r) + \frac{3
\alpha_{02}}{\alpha_0} u_3 (r) \nonumber \\ w_{0,\alpha} (r) &=& w_1
(r) - \frac{1}{\alpha_0} w_2 (r) + \frac{3 \alpha_{02}}{\alpha_0} w_3
(r) \nonumber \\ u_\beta (r) &=& \frac{1}{\alpha_0} u_2 (r) + \frac{3
\alpha_{2}}{\alpha_{02}} u_3 (r) -\frac1{15 \alpha_{02}} u_4 (r)
\nonumber \\ w_\beta (r) &=& \frac{1}{\alpha_0} w_2 (r) + \frac{3
\alpha_{2}}{\alpha_{02}} w_3 (r) -\frac1{15\alpha_{02}} w_4 (r)
\nonumber
\end{eqnarray}
where the functions $u_{1,2,3,4}$ and $w_{1,2,3,4}$ are independent on
$\alpha_0$, $\alpha_{02}$ and $\alpha_2$ and fulfill suitable boundary
conditions. As a consequence we get a linear correlation between $
1/\alpha_0$, $\alpha_{02}/\alpha_0$ and also a linear correlation
between $\alpha_2 /\alpha_{02}$ and $1/\alpha_{02}$. This means in
turn that according to the OPE potential both $\alpha_{02} $ and
$\alpha_2$ depends linearly with $ \alpha_{0}$. Numerically we get the
following correlations,
\begin{eqnarray}
\alpha_{02} &=& 0.963571370240 \, \alpha_0 - 3.467616391389 \nonumber \\ 
\alpha_{2}  &=& 3.467616391389 \, \frac{\alpha_{02}}{\alpha_0} + 5.080264230656
\label{eq:alpha_corr1}
\end{eqnarray} 
These relations are cut-off independent and unique consequences of the
OPE potential. On the other hand, the orthogonality between the bound
state and the scattering state yields
\begin{eqnarray} 
\alpha_0 &=& 1.037805911852 \,\alpha_{02} + 3.598712446758 \qquad \qquad (\alpha)
\nonumber \\ 
\alpha_{02} &=& 0.288382561043 \,\alpha_0\,\alpha_2 - 1.465059639612\,\alpha_0
\qquad (\beta)\nonumber\\
\label{eq:alpha_corr2}
\end{eqnarray}
The provided high accuracy is indeed needed. The four equations,
Eq.~(\ref{eq:alpha_corr1}) and Eq.~(\ref{eq:alpha_corr2}),
overdetermine the values of the three scattering lengths and could be
solved in triplets yielding four different solutions. Actually, there
are only two independent solutions which differences are compatible
within our numerical uncertainties. The scattering lengths and
effective range are presented in Table~\ref{tab:table_pert} and
compared to their perturbative value ( see Appendix~\ref{sec:pert} and
to the high quality Nijmegen potential
models~\cite{Stoks:1994wp}~\footnote{The values of $\alpha_0$ and
$r_0$ have been determined in Ref.~\cite{deSwart:1995ui}, whereas
$\alpha_{02}$ and $\alpha_2$ have been determined by us in
Ref.~\cite{PavonValderrama:2004nb}. See also
Ref.~\cite{PavonValderrama:2004se} for a extensive determination in
all partial waves.}. As we see, the agreement with the high quality
potentials is at the few percent level. Perturbation theory does not
account for most of the contribution to the effective range since the
orthogonality constraints preclude a short distance contribution to
$r_0$ and also to the deuteron matter radius $r_m$. This means in
practice that the counterterm named $C_2$ in
Refs.~\cite{Kaplan:1998sz,Kaplan:1998we} must vanish (See Appendix
\ref{sec:pert} for a detailed discussion). The dependence of the
scattering lengths $\alpha_0$, $\alpha_{02}$ and $\alpha_2$ on the
deuteron wavenumber $\gamma$ can be seen in Fig.~\ref{fig:a0[Gamma]},
where $\gamma$ dependent generalizations of the correlations,
Eq.~(\ref{eq:alpha_corr1}) and Eq.~(\ref{eq:alpha_corr2}) hold.

The phase shifts look very similar to previous
work~\cite{Beane:2001bc} using an energy expansion of a square well
potential as a counter-term and adjusting the depth of the two lowest
orders to reproduce the $^3S_1$ scattering length $\alpha_0 $ and
effective range $r_0$ as independent parameters or our variable phase
approach with non-trivial initial conditions in
Ref.~\cite{PavonValderrama:2004nb} where the full coupled channel
S-matrix was tailored to reproduce the effective range expansion to
any order treating all parameters as independent. Roughly speaking,
both approaches could be mapped to an energy dependent boundary
condition with no {\it a priori} orthogonality
constraints~\footnote{The main difference in this regard has to do
with the multi-valuation problem of the potential counter-term in
Ref.~\cite{Beane:2001bc} typical of inverse scattering problems. The
approach of Ref.~\cite{PavonValderrama:2004nb} does not have this
problem.}. The fact that the orthogonality constrained boundary
condition generates the bulk of the low energy threshold parameters
with only {\it one parameter} naturally explains the similarity
between the present phase shifts and those in previous
works~\cite{Beane:2001bc} and suggests that there is perhaps no need
to make the short distance boundary condition energy dependent if the
short distance cut-off is removed. 

\section{Short distance solutions and determination of the coefficients}
\label{sec:short} 

In this section we determine the coefficients of the OPE deuteron wave
functions appearing at short distances in Eq.~(\ref{eq:short_bc}). In
particular, we compute the energy independent and OPE potential
parameters independent short distance phase $ \varphi $. Let us remind
that any choice of $\varphi$ corresponds to a different choice of
short distance physics; given $\varphi$ and the OPE potential all
deuteron and scattering properties are uniquely determined. However,
the leading asymptotic form cannot directly be used to match the
numerical solution obtained by integrating in the large distance
solution. On the one hand, if we use cut-off approaches to determine
the regular solution, there are short distance cut-off effects when
the distance gets close to the cut-off radius. On the other hand, the
fact that the diverging exponential dominates over the converging one
provides too weak a signal for the corresponding coefficient. To
remedy the situation we improve on the short distance solution to
provide a reliable approximation at larger distances ($\sim 1{\rm
fm}$) where the diverging exponential is less dominant, and look for
plateaus in the matching radius. It turns out (see below) that one
should go at eight order in this expansion for a robust determination
of the short distance coefficients. Actually, one can then directly
match the short distance improved wave functions to the numerical
solution without no reference to cut-offs. We will try the two methods
and see that they yield to compatible results for the short distance
coefficients.

\begin{figure}[]
\begin{center}
\epsfig{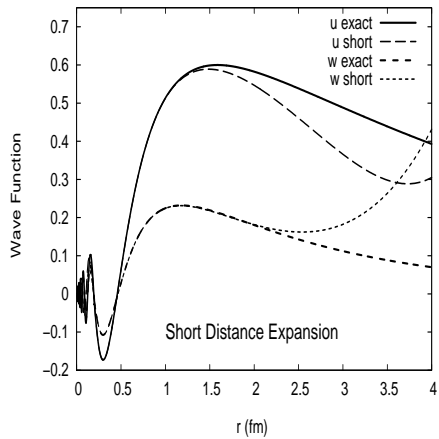}
\end{center}
\caption{The short distance expansion for the deuteron wave functions
matched to the numerical solution at a distance about $ r= 1 {\rm
fm}$. We take $ C_{1A}=0.1328 $, $C_{2A} = 0.2277 $, $C_{2 R}= -0.46 $
and $C_{1R} =0$. The numerical solutions are normalized by taking $A_S = 1$.
}
\label{fig:u+w_short.epsi}
\end{figure}

In the limit $r\to 0 $ the solutions to the coupled equations can be
written in an expansion of the form~\footnote{This expansion looks
similar to a coupled channel WKB expansion but it is free of some
inconveniencies. The applicability condition of the coupled channel
WKB method would be that the de Broglie local wavelength {\it matrix}
should be a slowly varying function of distance, implying in turn
three conditions on the corresponding local wavelength eigenvalues as
well as the corresponding WKB mixing angle which need not be
necessarily satisfied {\it simultaneously}, generating conversion mode
problems.}
\begin{eqnarray}
u(r) &=& u_0 \left(\frac{r}{R}\right)^{a_1}  e^{a_0
\sqrt{\frac{R}{r}}} f(r) \nonumber \\ 
w(r) &=& w_0
\left(\frac{r}{R}\right)^{a_2}  e^{a_0
\sqrt{\frac{R}{r}}} g(r) 
 \nonumber \\
\end{eqnarray} 
with 
\begin{eqnarray}
f(r) &=& \sum_{n=0}^\infty A_n
\left(\frac{r}{R}\right)^{n/2} \nonumber \\ 
g(r) &=& \sum_{n=0}^\infty B_n
\left(\frac{r}{R}\right)^{n/2} \nonumber \\
\end{eqnarray} 
At leading order we get the equations 
\begin{eqnarray}
u_0 a_0^2 + 16 \sqrt{2} w_0 &=&0 \nonumber \\
16 \sqrt{2} u_0 + (a_0^2 - 16)  w_0 &=&0 
\label{eq:short_LO}
\end{eqnarray} 
which have the four non trivial solutions, 
\begin{eqnarray}
(1A) \, , \quad a_0 &=& - 4 i \, , \qquad w_0 = \frac{u_0}{\sqrt{2}} \, , \\
(2A) \, , \quad a_0 &=& + 4 i \, , \qquad w_0 = \frac{u_0}{\sqrt{2}} \, , \\
(2R) \, , \quad a_0 &=& - 4 \sqrt{2} \, , \qquad w_0 = -\sqrt{2} u_0 \, , \\
(1R) \, , \quad a_0 &=& + 4  \sqrt{2} \, , \qquad w_0 = -\sqrt{2} u_0 \, .  
\end{eqnarray}  
The next to leading order equation becomes compatible only if 
\begin{eqnarray}
a_1 = a_2 = 3/4  
\end{eqnarray} 
For any solution in Eq.~(\ref{eq:short_LO}) we may then solve for the
remaining coefficients. One peculiar feature of this expansion is that
if one wants to determine the solution to a given order, one has to
compute the coefficients at a higher order. The reason is that strictly
speaking a truncation of the expansion involves also non-diagonal
elements, and one has the freedom to choose between solving $u$ or
$w$ to a given accuracy. The explicit result to eight order is
presented in Appendix~\ref{sec:short_app}. The general short distance
solution is written as a linear combination of the four independent
solutions,
\begin{eqnarray}
u (r) &=& \frac1{\sqrt{3}}\left(\frac{r}{R}\right)^{3/4} \Big[ -C_{1R}
  f_{1R} (r) e^{+ 4 \sqrt{2} \sqrt{\frac{ R}{r}}} \nonumber \\ &-&
  C_{2R} f_{2R}(r) e^{- 4 \sqrt{2} \sqrt{\frac{ R}{r}}} + \sqrt{2}
  C_{1A} f_{1A}(r) e^{- 4 i \sqrt{\frac{ R}{r}}} \nonumber \\ &+&
  \sqrt{2} C_{2A} f_{2A}(r) e^{ 4 i\sqrt{\frac{ R}{r}}} \Big]
  \nonumber \\ w (r) &=& \frac1{\sqrt{3}}
  \left(\frac{r}{R}\right)^{3/4} \Big[ \sqrt{2} C_{1R} g_{1R} (r) e^{+
  4 \sqrt{2} \sqrt{\frac{ R}{r}}} \nonumber \\ &+& \sqrt{2} C_{2R}
  g_{2R}(r) e^{- 4 \sqrt{2} \sqrt{\frac{ R}{r}}} + C_{1A} g_{1A}(r)
  e^{- 4 i \sqrt{\frac{ R}{r}}} \nonumber \\ &+& C_{2A} g_{2A}(r) e^{
  4 i\sqrt{\frac{ R}{r}}} \Big] \nonumber \\
\end{eqnarray} 
This expansion converges rather fast for each solution up to distances
of about $r \sim 0.6-0.9 {\rm fm}$. That is about what one needs,
since that is sufficiently far above the cut-off radius $ a\sim 0.1
{\rm fm}$.  Matching $u,w,u'$ and $w'$ at some point in this region we
get a linear relation between $C_{1R},C_{2R},C_{1A},C_{2A}$ and
$\eta$. Actually, we find that the signal of the converging
exponential is about hundred to thousand times that of the diverging
exponential in the range between 0.6 and 1 fm~\footnote{For instance
at $ r= 0.8 {\rm fm}$ we get
\begin{eqnarray} 
 u &=& 1.9683 - 59.4526\,\eta \nonumber \\ &=& 1.14054\, \bar
 C_{1A} - 52.3866\,C_{1R} + 1.09959\, \bar C_{2A} -
 0.00169556\,C_{2R}, \nonumber \\ w &=& -4.00531 + 159.28\,\eta
 \nonumber \\ &=& -0.667631\, \bar C_{1A} - 551.52\, C_{1R} +
 1.67549\, \bar C_{2A} + 0.0177228\, C_{2R} \nonumber \\ u' &=&
 -6.84126 + 287.992\,\eta \nonumber \\ &=& -1.34949\, \bar C_{1a} +
 273.671\, C_{1R} + 4.02872\, \bar C_{2A} - 0.00925419\, C_{2R}
 \nonumber \\ w' &=& 16.2726 - 607.14\,\eta \nonumber \\ &=&
 -0.667631\, \bar C_{1A} - 551.52\, C_{1R} + 1.67549\, \bar C_{2A} +
 0.0177228\,C_{2R}  \nonumber 
\end{eqnarray} 
where the l.h.s. corresponds to the numerical solution and the
r.h.s. to the short distance approximation, and the barred
coefficients $\bar C_{1A} = (C_{1A}+ C_{2A})/2$ and $\bar C_{2A} =
(-C_{1A} + C_{2A})/2\,i$ have been introduced.}.  Matching directly
the integrated in solution to the short distance solution with a
vanishing coefficient of the diverging exponential $C_{1R}=0$, we get
at the scale $ 0.7 < r < 0.9 {\rm fm} $
\begin{eqnarray}
C_{2R}&=& -0.47(1) \, , \quad \eta = 0.0263333(1) \nonumber \\ 
\bar C_{1A} &=& 0.1327(3)  \, \quad , \bar C_{2A} = 0.2277(5) 
\end{eqnarray} 
We can instead determine $\eta=0.0263332 $ from the boundary condition
BC6 in Eq.~(\ref{eq:bc_a}) at $r=0.2 {\rm fm} $ and deduce the
remaining constants yielding 
\begin{eqnarray}
|C_{1R}| &<&  10^{-7} \, ,  \quad  C_{2R} = -0.47 (1) \nonumber \\ 
\bar C_{1A} &=& 0.1327(3)\, ,  \quad \bar C_{2A} = 0.2277 (5)  
\end{eqnarray} 
The errors have been estimated by varying the matching point in the
region $0. 7 {\rm fm } < r < 0.9 {\rm fm } $. Note that although the
coefficient of the diverging exponential $C_{1R}$ is six orders of
magnitude larger than the one of the converging exponential, the
solution through the matching condition is about eight orders of
magnitude smaller (may even change sign ).  So that the result
provides a sizeable signal for the converging exponential.  With these
values we show in Fig.~\ref{fig:u+w_short.epsi} the short distance
wave functions compared to the integrated in numerical ones when the
matching is undertaken at $ r=0.8 {\rm fm}$. To improve on the short
distance side we have taken $C_{1R}=0$. The error in the region $0. 7
{\rm fm } < r < 0.9 {\rm } $ never exceeds a $0.01\%$.  We have
checked that setting the constant $C_{2R}=0$ introduces a larger
deviation from the numerical solution as compared to the computed
value in the region above $1 {\rm fm}$. Finally, the corresponding
short distance angle reads
\begin{eqnarray}
\varphi = - \tan^{-1} \frac{\bar C_{2A}}{\bar C_{1A}}= -59.7 (1)^o \, . 
\end{eqnarray} 
The discussion in this section explicitly shows that contrary to the
findings in Ref.~\cite{Martorell94} the coefficient of the converging
exponential does not vanish.

\section{Conclusions and Outlook} 
\label{sec:concl}

In this paper we have reanalyzed the OPE potential in the triplet
$^3S_1-{}^3D_1$ channel both for bound and scattering states. Rather
than modeling the interaction below some finite short distance we have
adopted the viewpoint of taking the potential seriously down to the
origin. This must be carefully done and in a way as to get rid of any
short distance ambiguities. In addition, this procedure proves crucial
to be able to disentangle the OPE contribution from other
contributions, like TPE and higher, electromagnetic effects and
relativistic corrections to deuteron and NN scattering
observables. Our analysis is carried out entirely in coordinate space
where these corrections generate a potentical which is finite
everywhere except at the origin. Momentum space treatments require an
additional regularization of the potential. 

The OPE coupled channel potential is singular at short distances and
additional conditions need to be specified on the wave functions at
the origin. Actually, the singular eigen-potentials at short distances
are attractive and repulsive and while in the attractive case a mixed
boundary condition specifies the corresponding short distance
eigenfunction, in the repulsive case one must impose a standard
homogeneous boundary condition. This only leaves one free parameter,
which we have chosen to be the deuteron binding energy and which
cannot be determined from the OPE potential. All remaining deuteron
observables come out for free. For the scattering states in the
$^3S_1-{}^3D_1$ channel, we have demanded orthogonality constraints
between all states of different energy. This condition is actually an
additional requirement for singular potentials, since the
orthogonality relation carries information on the peculiar short
distance behavior of the wave functions, and is not necessarily
satisfied. The most obvious example where orthogonality constraints
are violated corresponds to energy dependent potentials and energy
dependent boundary conditions in coordinate space. A less trivial but
significant example is the case of dimensional regularization in the
PDS scheme as a perturbative analysis in coordinate space of both bound
and scattering states reveals. The power of the orthogonality constraints
for singular potentials is that all scattering properties are then
predicted from the OPE potential parameters and the deuteron binding
energy. 

In our analysis it turns out that the short distance form of all wave
functions is characterized by some short distance constants. We have
clarified the role played by the exponentially suppressed regular
solution by determining its non-vanishing value numerically using
short distance expansions to high order, to explore the region below
$0.1 {\rm fm} $, not accessible to standard numerical integration
methods. Another relevant constant is given by a short distance phase
$\varphi$ which plays the role of a fundamental dimensionless constant
in the OPE problem.  It does not depend on the energy nor on the OPE
parameters, but it is related to the form of the OPE potential in the
chiral limit. The closeness of this phase to $\pi /3 $ is mysterious
and suggestive and requires further investigation.

It is remarkable that indeed the bulk of the experimental results both
for the bound state as well the scattering observables are accounted
for at the $2-3 \%$ level by the OPE potential taken from zero to
infinity. We interpret this success as a confirmation on the validity
of our choice of regular solutions and the use of orthogonality
constraints. The discrepancies can legitimately be attributed to other
effects such as TPE, electromagnetic and relativistic
corrections. Many of the methods and results obtained in this paper
can be generalized in a straightforward manner to take these effects
into account and to the study of higher partial waves without any
substantial modifications. In particular, the number of independent
constants in a given channel depends on the short distance behavior of
the long range potential. The bonus of such a program would be the
complete elimination of short distance ambiguities in the study of the
NN interaction with known long distance forces as determined by chiral
symmetry. In our view this an indispensable prerequisite to asses the
relevance of chiral symmetry in nuclear physics in a model independent
way. The systematic study of these effects will be reported
elsewhere~\cite{Pavon_TPE}.

\begin{acknowledgments}

We thank J. Nieves for a critical reading of the manuscript.  This
research was supported by DGI and FEDER funds, under contract
BFM2002-03218 and by the Junta de Andaluc\'\i a.

\end{acknowledgments}

\appendix

\section{Perturbative solutions}
\label{sec:pert}

\subsection{Bound state} 

\begin{figure}[]
\begin{center}
\epsfig{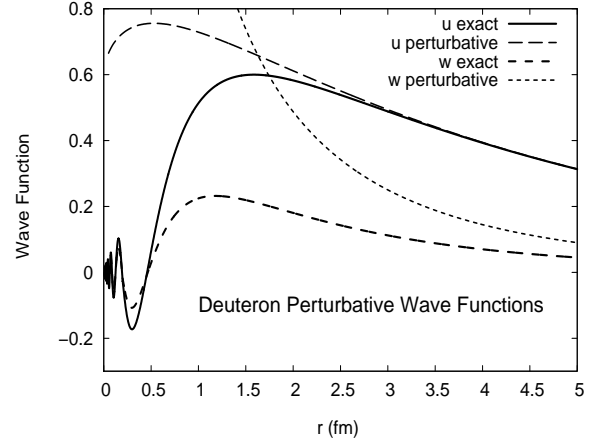}
\end{center}
\caption{Perturbative deuteron wave functions compared to the exact
ones as a function of the distance (in {\rm fm} ).
The exact ones are normalized by taking $A_S = 1$.
}
\label{fig:u+w_perturbative}
\end{figure}

In this appendix we solve the coupled deuteron equations,
Eq.~(\ref{eq:sch_coupled}) in standard perturbation theory for the
fixed negative energy bound state. A somewhat related
approach looking for the equivalence with the PDS scheme
of~\cite{Kaplan:1998we} in the one-channel positive energy case can be
looked up in Ref.~\cite{Cohen:1998bv}. The problem of orthogonality
was not discussed.  The requirement of normalizability of the
deuteron state requires the D wave component to vanish. Thus, at
lowest order we have the normalizable solutions,
\begin{eqnarray} 
u_\gamma^{(0)} (r) &=& e^{-\gamma r} \nonumber \\ 
w _\gamma^{(0)}(r) &=&0
\end{eqnarray} 
At first order we have to solve the equations
\begin{eqnarray}
-u_\gamma^{(1) \prime \prime }  (r) + \gamma^2 u_\gamma^{(1)} (r) &=& - U_s (r)
e^{-\gamma r} \, ,\nonumber \\ -w_\gamma^{(1) \prime \prime }  (r) +
\left[\frac{6}{r^2} + \gamma^2 \right] w_\gamma^{(1)} (r) &=&
-U_{sd}(r) e^{-\gamma r} \, , \nonumber \\
\label{eq:sch_coupled_pert} 
\end{eqnarray}
Using the regular and irregular solutions at the origin 
\begin{eqnarray}
u_{\rm reg}( r) &=& 2 \frac{\sinh ( \gamma r )}{\gamma r} \nonumber \\
w_{\rm reg}( r) &=& 2 \left( 1+ \frac{3}{(\gamma r)^2} \right) \sinh (
\gamma r ) - \frac{6}{\gamma r} \cosh( \gamma r ) \nonumber \\
u_{\rm irreg} (r) &= & e^{-\gamma r} \, . \\ w_{\rm irreg} (r) & = & \
e^{-\gamma r} \left( 1 + \frac{3}{\gamma r} + \frac{3}{(\gamma r)^2} \right)
\, .
\end{eqnarray}
we get 
\begin{eqnarray} 
u_\gamma^{(1)} (r) &=& \int_0^\infty G_s (r,r') U_s (r') e^{-\gamma
r'} dr' \\ w_\gamma^{(1)} (r) &=& \int_0^\infty G_d (r,r') U_{sd}
(r') e^{-\gamma r'} dr'
\end{eqnarray} 
where $G_s$ and $G_d$ are the corresponding Green functions. 
Explicit calculation yields 
\begin{widetext} 
\begin{eqnarray}
u_\gamma^{(1)} (r) &=& e^{-\gamma r} \frac{m^2\,R\,
        \Gamma(0,m\,r + 2\,r\,\gamma )-\Gamma(0,m\,r)}{3\,\gamma } - \frac{2\,m^2\,R\,
     {\rm Ei}(- m\,r   - 2\,r\,\gamma )\,
     \sinh (r\,\gamma )}{3\,\gamma } \\ 
w_\gamma^{(1)} (r) &=& {e^{-r\,\gamma }} \left( 1 + \frac{3}{r^2\,{\gamma
}^2} + \frac{3}{r\,\gamma } \ \right) \times \nonumber \\ &&
\,\Big[\frac{m^2\,R\, \left( 3\,m^2 - 4\,{\gamma }^2 \right) \, 
\Gamma(0,m\,r + 2\,r\,\gamma )
-\Gamma(0,m\,r )}{6\,{\sqrt{2}}\,{\gamma }^3} \nonumber \\
&+& \frac{R\, \left( -6\,m^3\,\gamma + 6\,m^2\,{\gamma }^2 +
4\,{\gamma }^4 + \left( 3\,m^4 - 4\,m^2\,{\gamma }^2 \right) \, \log
(1 + \frac{2\,\gamma }{m}) \right) }{6\,{\sqrt{2}}\, {\gamma }^3}
\nonumber \\ &+& \frac{e^{-m\,r - 2\,r\,\gamma }\,R\, \left( 6 +
6\,m\,r + m^2\,r^2 - m^3\,r^3 + 4\,r\,\gamma + 4\,m\,r^2\,\gamma +
2\,m^2\,r^3\,\gamma \right) }{2\, {\sqrt{2}}\,r^4\,{\gamma }^3}
\nonumber \\ &-& \frac{R\, \,e^{-m\,r} \left( 6 + 6\,m\,r + m^2\,r^2 -
m^3\,r^3 - 8\,r\,\gamma - 8\,m\,r^2\,\gamma + 4\,r^2\,{\gamma }^2 +
4\,m\,r^3\,{\gamma }^2 \right) }{2\,{\sqrt{2}}\,r^4\, {\gamma }^3}
\nonumber \Big] \nonumber \\ &+& \left( 2 \left( 1+ \frac{3}{\gamma^2
r^2} \right) \sinh ( \gamma r ) - \frac{6}{\gamma r} \cosh( \gamma r
)\right) \times \nonumber \\ && \Big[ \frac{e^{- m\,r - 2\,r\,\gamma
}\,R\, \left( 6 + 6\,m\,r + m^2\,r^2 - m^3\,r^3 + 4\,r\,\gamma +
4\,m\,r^2\,\gamma + 2\,m^2\,r^3\,\gamma \right) }{2\,{\sqrt{2}}\,
r^4\,{\gamma }^3} \nonumber \\ &-& \frac{m^2\,R\, \left( 3\,m^2 -
4\,{\gamma }^2 \right) \, {\rm Ei}(- m\,r - 2\,r\,\gamma
)}{6\, {\sqrt{2}}\,{\gamma }^3} \Big]
\end{eqnarray} 
\end{widetext} 
where $\Gamma(0,z)$ and ${\rm Ei} (z) $ are the standard incomplete
Gamma function and the Exponential integral function respectively 
\begin{eqnarray}
\Gamma(0,z) &=& \int_z^\infty dt \frac{e^{-t}}{t} \\ 
{\rm Ei} (z) &=& - P\int_{-z}^\infty dt \frac{e^{-t}}{t} 
\end{eqnarray} 
At asymptotically large distances we have  
\begin{eqnarray}
u_\gamma^{(1)} (r) &\to & c_{\rm pert} e^{-\gamma r} \\ w_\gamma^{(1)} (r)
&\to & \eta_{\rm pert} e^{-\gamma r} \left( 1 + \frac{3}{\gamma r} +
\frac{3}{(\gamma r)^2} \right)
\end{eqnarray} 
where 
\begin{eqnarray}
c_{\rm pert} &=& \int_0^\infty U_{s} (r) u_{\rm reg} (r)  e^{-\gamma r} dr  \\ 
\eta_{\rm pert} &=& \int_0^\infty U_{sd} (r) w_{\rm reg} (r)  e^{-\gamma r} dr 
\end{eqnarray} 
Explicit calculation yields 
\begin{eqnarray}
c_{\rm pert} &=& \frac{ R m^2}{3 \gamma} \log \left( 1 +
\frac{2\gamma}m \right)\\ \eta_{\rm Pert} &=& \frac{R}{6\sqrt{2}m
\gamma^3} \Big[ 4 \gamma^4 + 6m^2 \gamma^2 - 6 m^3 \gamma \nonumber \\
&+& ( 3 m^4 - 4 m^2 \gamma^2 ) \log\left( 1 + \frac{2 \gamma}{m}
\right) \Big] \nonumber \\ &=& \frac{32 \sqrt{2} R}{45 m} \gamma^2 -
\frac{2 \sqrt{2} R\gamma^3}{3m^2} + \dots 
\label{eq:eta-pert} 
\end{eqnarray} 
The numerical value we get is $\eta_{\rm pert}=0.0510 $ almost twice
the exact OPE result. Taking this perturbative value for $\eta $ we
show in Fig.~\ref{fig:u+w_perturbative} the perturbative deuteron wave
functions as compared to the exact ones.

Unfortunately, if one wants to improve on this first order
calculation going to second order perturbation theory there is a problem
since the behavior of the perturbative wave functions at short
distances is given by
\begin{eqnarray}
u_\gamma^{(1)} (r) &=& -\frac{ R m^2}{3 \gamma} \log \left( 1 +
\frac{2\gamma}m \right) + \dots \\ w_\gamma^{(1)} (r) &=&
\frac{\sqrt{2}R}{r} - \frac23 \sqrt{2} R \gamma + \dots
\end{eqnarray} 
making the wave function non normalizable, unlike the exact regular
wave function. This divergence at short distances actually precludes
going to higher orders in perturbation theory.
\medskip

The normalization at first order is given by 
\begin{eqnarray}
\frac1{A_S^2} = \int_0^\infty ( e^{-2 \gamma r} + 2 u_\gamma^{(1)} (r)
e^{-\gamma r} )
\end{eqnarray} 
and hence 
\begin{eqnarray}
\frac{A_S}{\sqrt{2 \gamma}}  &=& 1 - \frac{ 2R m^2 }{3(m + 2 \gamma)}
+\frac{ R m^2}{3 \gamma} \log \left( 1 + \frac{2\gamma}m \right) \nonumber \\ &=& 1 - \frac{ 2R
\gamma}3 - \frac{16 R \gamma^2}{9m} + \dots    
\end{eqnarray} 
The deuteron matter radius is given by 
\begin{eqnarray}
r_{m,\,\rm pert}^2 = \frac14 A_S^2 \int_0^\infty r^2 ( e^{-2 \gamma r} + 2
u_\gamma^{(1)} (r) e^{-\gamma r} )
\end{eqnarray} 
and hence to first order one has 
\begin{eqnarray}
r_{m,\,\rm pert}^2 = \frac{1}{8 \gamma^2} + \frac{m^2 R ( 3 m +10 \gamma)}{18\gamma (m + 2 \gamma)^3 } + \dots 
\end{eqnarray} 
yielding in the weak binding regime
\begin{eqnarray} 
\sqrt{8} \gamma \,\, r_{m,\,\rm pert} &=& 1 + \frac{2 R \gamma}{3} - \frac{16
R \gamma^2}{9 m} + \dots   
\end{eqnarray} 
Finally, the quadrupole moment at first order is given by
\begin{eqnarray}
Q_{\rm pert}= \frac{\sqrt{2}}{10} \int_0^\infty r^2 w_\gamma^{(1)} (r)
e^{-\gamma r} dr
\end{eqnarray} 
The integral can be evaluated to give
\begin{eqnarray}
Q_{\rm pert} &=& \frac{8 R ( 4 m^2 + 9 \gamma m + 6 \gamma^2 )}{45 (m
+ 2 \gamma )^3} \nonumber \\ 
&=& \frac{32 R}{45 m} - \frac{8 R \gamma}{3m^2} + \frac{128R \gamma^2}{15 m^3} + \dots   
\end{eqnarray}
\phantom{bla}\\ 
yielding $ Q_{\rm pert}= 0.4555 {\rm fm}^2  $.

Our perturbative expressions for $A_S$, $r_m$ and $Q$ coincide with
those of Kaplan Savage and Wise~\cite{Kaplan:1998sz} provided one
takes in their expression for $r_m$ the renormalization scale in the
PDS scheme to be $\mu = \gamma $, instead of taking $\mu = m$ as they
do or else taking the $C_2$ counter-term identically equal to
zero. Actually, the $C_2$ counter-term can be mapped into a short
distance contribution to the effective range parameter $r_0$ in the
$^3S_1$ channel. The value of $\eta$ was not given in that reference
but can be deduced from the off-diagonal scattering amplitude in the
$^3S_1-{}^3D_1 $ channel given in their previous
work~\cite{Kaplan:1998we} by evaluating the residue at the deuteron
pole. The result also agrees with the calculation presented here.

\subsection{Low energy parameters} 

To check the identification $C_2=0$ further let us compute the S-wave
effective range $r_0$. For our purposes it is sufficient to analyze
the zero energy scattering state. The lowest order solution is given by an
$\alpha $ state 
\begin{eqnarray}
u_{0,\alpha}^{(0)} (r) &=& \left( 1 - \frac{r}{\alpha_0} \right) 
 \nonumber \\
w_{0,\alpha}^{(0)} (r) &=& 0 
\end{eqnarray} 
At zeroth order in the OPE coupling the orthogonality constraint
yields
\begin{eqnarray}
0 &=& \int_0^\infty u_{0,\alpha}^{(0)} (r) u_\gamma^{(0)} (r) dr \nonumber \\
&=& \frac{1}{\gamma^2 } \left[-\frac1{\alpha_0} + \gamma
\right]
\end{eqnarray} 
which yields the scattering length to lowest order,
\begin{eqnarray}
\alpha_0^{(0)} = \frac1{\gamma} 
\end{eqnarray}  
At first order we use the regular solution $u_{\rm reg} (r)= r $ and
the irregular solution $u_{\rm irreg} (r) = (1-r/\alpha_0) $ and get
similarly to the bound state case the first order correction to the
$\alpha $ state,
\begin{eqnarray}
u_{0,\alpha}^{(1)} (r) &=& \frac{2 R }{3\, \alpha_0} e^{-m\,r} \left(1
- \alpha_0 \,m \right) - \frac{2\,m^2\,r\,R}{3} {\rm Ei}(- m\,r )
\nonumber \\ 
w_{0,\alpha}^{(1)} (r) &=&
\frac{\,e^{-m\,r}\,R}{15\,{\sqrt{2}}\,\alpha_0 \,m^2\,r^2} \nonumber \\ 
&\times& \Big( 120
-64\,\alpha_0 \,m + 120\,m\,r - 34\,\alpha_0 \,m^2\,r \nonumber \\ &+& 40\,m^2\,r^2 -
2\,\alpha_0 \,m^3\,r^2  + \alpha_0 \,m^4\,r^3 - \alpha_0 \,m^5\,r^4
\Big) \nonumber \\ &+&  \frac{R\,\left( -120 + 64\,\alpha_0 \,m +
\alpha_0 \,m^6\,r^5\,\Gamma(0,m\,r) \right) } {15\,{\sqrt{2}}\,\alpha_0
\,m^2\,r^2} \nonumber \\ 
\end{eqnarray}  
Note that asymptotically the first order correction to the S-wave
vanishes exponentially and hence cannot contribute to the scattering
length. On the other hand, the orthogonality relation to first order
reads
\begin{eqnarray} 
0 = \int_0^\infty dr \left[ u_{\gamma}^{(0)} u_{0,\alpha}^{(0)} +
u_{\gamma}^{(1)} u_{0,\alpha}^{(0)} + u_{\gamma}^{(1)}
u_{0,\alpha}^{(0)} \right] 
\end{eqnarray}  
and after computing the integrals one gets 
\begin{eqnarray} 
0= - \frac1{\alpha_0}+ \gamma &+& \frac{ R }{3 \alpha_0 \gamma} \Big[
m^2 ( 1 + \alpha_0 \gamma ) \log\left( 1 + \frac{2 \gamma}{m} \right)
\nonumber \\ &-& 2\gamma ( m - \gamma + \alpha_0 m \gamma ) \Big]
\end{eqnarray} 
Solving perturbatively for $\alpha_0$ we get  at first order 
\begin{eqnarray}
\alpha_{0,{\rm pert}} = \frac{1}{\gamma} - \frac{ 2 m^2 R }{3
\gamma^2} \left[ \frac{\gamma ( \gamma- 2 m ) }{m^2} + \log\left( 1 +
\frac{2 \gamma}{m} \right) \right] + \dots \nonumber \\
\end{eqnarray}  
Numerically one gets $ \alpha_{0,{\rm pert}}= (4.3177 + 0.2912 +
\dots) {\rm fm} $ to be compared with the full OPE result $\alpha_0 =
5.34 $ and the experimental value $ \alpha_0 = 5.42 {\rm fm} $. The
$E_1$ scattering length $\alpha_{02}$ can be read off from the D-wave,
using the asymptotic condition in Eq.~(\ref{eq:zero_energy})
\begin{eqnarray}
\alpha_{02} &=& \frac{4 \sqrt{2} R (15 - 8 \alpha_0 m ) }{45 m^2}
\nonumber \\ &=& \frac{4 \sqrt{2} R (15 \gamma - 8 m ) }{45 \gamma
m^2}
\end{eqnarray}  
in the second line we have substituted the perturbative relation
$\alpha_0 = 1/ \gamma + {\cal O} (R) $. Note the linear correlation
$\alpha_{02} = 1.5499 \alpha_0- 4.1530 $ to be compared with the exact
OPE relation in Eq.~(\ref{eq:alpha_corr1}). The numerical value one
gets for the first and second lines taking the experimental values of
$\alpha_0 = 5.42 $ and $\gamma $ are $\alpha_{02} = 4.24 {\rm fm^3}$
and $\alpha_{02} = 2.53 {\rm fm^3}$ respectively to be compared with
the experimental $\alpha_{02}=1.64 {\rm fm^3}$.  In the weak binding
limit one obtains
\begin{eqnarray}
\gamma \,\alpha_{0,{\rm pert}} = 1 + \frac{2R\gamma}3 - \frac{16 R
\gamma^2}{9 m} + \dots
\end{eqnarray}  
In this limit we have the perturbative linear correlation between the
scattering length and the deuteron matter radius
\begin{eqnarray} 
r_m = \frac{\alpha_0}{2\sqrt{2}} + {\cal O} (\gamma^3 , R^2)  
\end{eqnarray} 
which yields the value $r_m = 1.92 $ for the experimental scattering
length $\alpha_0 = 5.42 {\rm fm} $. The linear correlation was
established empirically with realistic potentials in
Ref.~\cite{Martorell:1986,Martorell:1990}.

To first order the effective range in the $^3S_1$ eigen-channel is
given by
\begin{eqnarray} 
r_0 = - 4 \int_0^\infty dr \,  u_{0,\alpha}^{(0)}  (r) u_{0,\alpha}^{(1)}  (r) 
\end{eqnarray} 
yielding 
\begin{eqnarray}
r_{0,{\rm pert}} &=& \frac{4R(3m^2- 8 \gamma m + 6 \gamma^2 ) }{9m^2}
\nonumber \\ &=& 1.4369 - 5.4789 \gamma + 5.8758 \gamma^2 \nonumber \\ 
&=& 0.4831
          {\rm fm} 
\end{eqnarray}  
a result much smaller than the full OPE result ($ 1.64\,{\rm fm} $) and
the experimental number ($1.75\,{\rm fm} $). Again, our result
corresponds to a theory where the short distance contribution to the
effective range vanishes, i.e. $C_2=0$. A non vanishing value of $C_2$
was needed to fit the experimental values of both the matter radius
and the effective range.  Our calculation shows that the scheme
developed in Refs.~\cite{Kaplan:1998sz} and Ref.~\cite{Kaplan:1998we}
does not fulfill perturbatively the orthogonality constraints.

\section{Short distance expansion}
\label{sec:short_app} 

For the $f(r)$ function we get (we use $ x=r/R$), \def\imag{i}
\begin{widetext}
\begin{eqnarray}
f_{1A}&=& 1 - \frac{35\,\imag }{32}\,{\sqrt{x}} - \frac{1811\,x}{6144}
  + \frac{2441\,\imag }{65536}\,x^{\frac{3}{2}} -
  \frac{34805\,x^2}{8388608}  \nonumber \\ &+& x^3\,\left(
  \frac{9873675}{17179869184} + \frac{m^2\,R^2}{36} -
  \frac{m^3\,R^3}{32} - \frac{3\,R^2\,{\gamma }^2}{64} \right)
  \nonumber \\ &+& x^{\frac{7}{2}}\,\left( \frac{193405905\,\imag
  }{549755813888} + \frac{353\,\imag }{24192}\,m^2\,R^2 -
  \frac{709\,\imag }{92160}\,m^3\,R^3 + \frac{\imag }{28}\,m^4\,R^4 -
  \frac{709\,\imag }{61440}\,R^2\,{\gamma }^2 \right) \nonumber \\ &+& 
  x^{\frac{5}{2}}\,\left( \frac{-333725\,\imag }{268435456} -
  \frac{\imag }{15}\,m^3\,R^3 - \frac{\imag }{10}\,R^2\,{\gamma }^2 \
  \right) \\ 
f_{2A}&=& f_{1A}^* \\ 
f_{2R}&=& 
1 + \frac{67\,{\sqrt{x}}}{32\,{\sqrt{2}}} + \frac{7763\,x}{12288} + 
  \left( \frac{8873}{131072\,{\sqrt{2}}} - 
     \frac{m^2\,R^2}{3\,{\sqrt{2}}} \right) \,x^{\frac{3}{2}} + 
  \left( -\frac{105845}{33554432}  - 
     \frac{55\,m^2\,R^2}{192} \right) \,x^2 \nonumber \\ &+& 
  \left( \frac{881405}{1073741824\,{\sqrt{2}}} - 
     \frac{10807\,m^2\,R^2}{184320\,{\sqrt{2}}} + 
     \frac{m^3\,R^3}{15\,{\sqrt{2}}} + 
     \frac{R^2\,{\gamma }^2}{10\,{\sqrt{2}}} \right) \,x^{\frac{5}{2}} \nonumber \\ &+&  
  \left( - \frac{23360715}{137438953472}   - 
     \frac{332899\,m^2\,R^2}{11796480} + \frac{47\,m^3\,R^3}{960} + 
     \frac{m^4\,R^4}{36} + \frac{47\,R^2\,{\gamma }^2}{640} \right) \,x^3 \nonumber \\ &+&  
  \left( \frac{419268465}{4398046511104\,{\sqrt{2}}} + 
     \frac{30559591\,m^2\,R^2}{31708938240\,{\sqrt{2}}} + 
     \frac{2141\,m^3\,R^3}{1290240\,{\sqrt{2}}} + 
     \frac{229\,m^4\,R^4}{8064\,{\sqrt{2}}} + 
     \frac{2141\,R^2\,{\gamma }^2}{860160\,{\sqrt{2}}} \right) \,
   x^{\frac{7}{2}} \nonumber \\
f_{1R} &=& f_{2R}   \qquad \left(  x \to e^{2\pi i} x \right) 
\end{eqnarray}
and for the $g(r)$ function one has 
\begin{eqnarray}
g_{1A}&=& 1 - \frac{35\,\imag }{32}\,{\sqrt{x}} - \frac{4883\,x}{6144}
  + \frac{82075\,\imag }{196608}\,x^{\frac{3}{2}} +
  \frac{1245195\,x^2}{8388608} \nonumber \\ &+& \left(
  \frac{-5136285\,\imag }{268435456} - \frac{\imag }{15}\,m^3\,R^3 -
  \frac{\imag }{10}\,R^2\,{\gamma }^2 \ \right) \,x^{\frac{5}{2}}
  \nonumber \\ &+& \left( \frac{42237195}{17179869184} -
  \frac{m^2\,R^2}{18} - \frac{m^3\,R^3}{32} - \frac{3\,R^2\,{\gamma
  }^2}{64} \right) \,x^3 \nonumber \\ &+& \left(
  \frac{494999505\,\imag }{549755813888} - \frac{65\,\imag
  }{12096}\,m^2\,R^2 + \frac{2363\,\imag }{92160}\,m^3\,R^3 +
  \frac{\imag }{28}\,m^4\,R^4 + \frac{2363\,\imag
  }{61440}\,R^2\,{\gamma }^2 \right) \,x^{\frac{7}{2}} +{\cal
  O}(x^{4}) \\ g_{2A} &=& g_{1A}^* \\ g_{2R}&=& 1 +
  \frac{67\,{\sqrt{x}}}{32\,{\sqrt{2}}} + \frac{13907\,x}{12288} +
  \left( \frac{307195}{393216\,{\sqrt{2}}} -
  \frac{m^2\,R^2}{3\,{\sqrt{2}}} \right) \,x^{\frac{3}{2}} + \left(
  \frac{5075595}{33554432} - \frac{55\,m^2\,R^2}{192} \right) \,x^2
  \nonumber \\ &+& \left( \frac{19661565}{1073741824\,{\sqrt{2}}} -
  \frac{41527\,m^2\,R^2}{184320\,{\sqrt{2}}} +
  \frac{m^3\,R^3}{15\,{\sqrt{2}}} + \frac{R^2\,{\gamma
  }^2}{10\,{\sqrt{2}}} \right) \,x^{\frac{5}{2}} \nonumber \\ &+&
  \left( - \frac{143137995}{137438953472} -
  \frac{128033\,m^2\,R^2}{3932160} + \frac{47\,m^3\,R^3}{960} +
  \frac{m^4\,R^4}{36} + \frac{47\,R^2\,{\gamma }^2}{640} \right) \,x^3
  \nonumber\\ &+ & \left( \frac{1476620145}{4398046511104\,{\sqrt{2}}}
  - \frac{45736601\,m^2\,R^2}{31708938240\,{\sqrt{2}}} +
  \frac{45149\,m^3\,R^3}{1290240\,{\sqrt{2}}} +
  \frac{229\,m^4\,R^4}{8064\,{\sqrt{2}}} + \frac{45149\,R^2\,{\gamma
  }^2}{860160\,{\sqrt{2}}} \right) \, x^{\frac{7}{2}} +{\cal O}(x^4)
  \nonumber \\ g_{1R} &=& g_{2R} \qquad \left( x \to e^{2\pi i} x
  \right)
\end{eqnarray} 
\end{widetext}

\section{LOCAL Diagonalization and perturbative mixing}
\label{sec:local_rot} 

One of the puzzles one encounters in the description of the deuteron
with the OPE potential is that while the dimensionless D/S ratio
parameter is rather small at long distances $w / u \to \eta =0.0256$,
it actually comes from a strong mixing at short distances where $ w/u
\to 1/\sqrt{2} \sim 0.707 $. Actually, the analog question for
scattering states is that there seems to be a natural hierarchy for
the phase shifts in the $^3S_1-{}^3D_1$ channel, namely $ \delta_{3S1}
\gg \delta_{3D1} \gg \epsilon_1 $ even though the threshold behavior
of the D-wave is more suppressed than that of the mixing angle. The
question is whether one can think of an expansion in terms of the
$\eta$ parameter. There are two obvious situations where the mixing
does not occur. One is the absence of tensor force. In the OPE
potential that would also eliminate the $D-wave$. Another situation is
dropping the mixing terms in the OPE potential, which is questionable
since they are actually larger than the diagonal terms. It is
possible, however, to write the equations in a form that the mixing is
manifestly small at {\it all} distances. To this end we make a local
rotation of the deuteron wave functions
\begin{eqnarray}
\begin{pmatrix}
u(r) \\ w(r)  
\end{pmatrix} = 
\begin{pmatrix}
\cos\theta(r) & \sin\theta(r)  \\ -\sin\theta(r) &  \cos \theta(r)   
\end{pmatrix} 
\begin{pmatrix}
u_A (r)\\ u_R (r) 
\end{pmatrix}  
\end{eqnarray} 
in such a way as to diagonalize the potential we have
\begin{eqnarray}
\begin{pmatrix}
U_s & U_{sd} \\ U_{sd} & U_{d} + \frac6{r^2}
\end{pmatrix}  
&&= \nonumber \\ 
\begin{pmatrix}
\cos\theta & \sin\theta  \\ -\sin\theta &  \cos \theta   
\end{pmatrix}  && 
\begin{pmatrix}
U_A  & 0  \\ 0 &  U_R    
\end{pmatrix}  
\begin{pmatrix}
\cos\theta & \sin\theta  \\ -\sin\theta &  \cos \theta   
\end{pmatrix} 
\end{eqnarray}  

\begin{figure*}[]
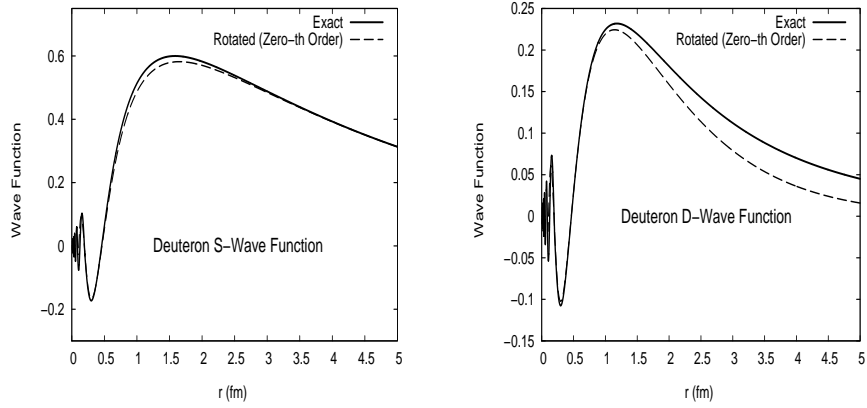

\begin{center}
\epsfig{figure=urotated.epsi,height=5.5cm,width=5.5cm}\qquad 
\epsfig{figure=wrotated.epsi,height=5.5cm,width=5.5cm}
\end{center}
\caption{The rotated eigenfunctions at zeroth order compared to the
exact ones. They correspond to take $ \eta=0$.}
\label{fig:urotated}
\end{figure*}

The deuteron equations for the OPE potentials read after the local rotation 
\begin{widetext}
\begin{eqnarray}
-u_A '' (r) + \left[ U_A (r) + \theta' (r)^2 \right] u_A (r) +
\gamma^2 u_A (r) &=& \left[2 \theta'(r) u_R' (r) + \theta''(r) u_R (r)
  \right] \, ,\nonumber \\ 
-u_R '' (r) + \left[ U_R (r) + \theta' (r)^2 \right] u_R (r) + \gamma^2 u_R  (r) &=&  -\left[2 \theta'(r) u_A' (r) + \theta''(r) u_A (r) \right] 
\, ,\nonumber  \\ 
\label{eq:sch_rotated} 
\end{eqnarray}
\end{widetext} 
In the coupled channel space these equations can be visualized as a
particle with spin in the presence of a gauge potential
$\theta'(r)$.
\medskip
\begin{figure*}[]
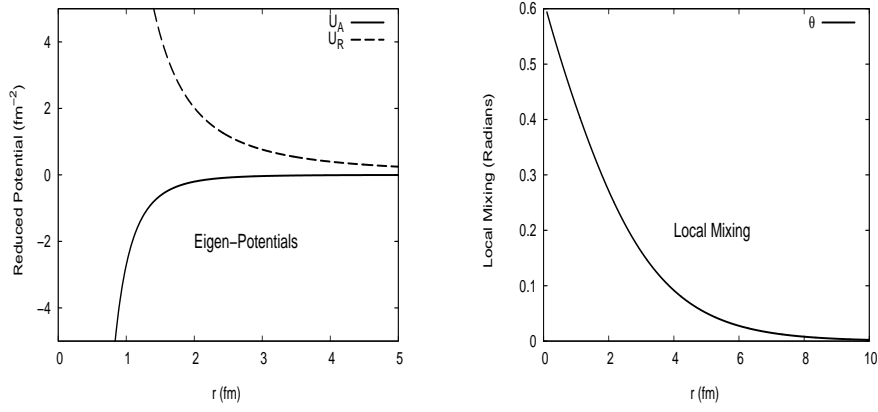

\begin{center}
\epsfig{figure=eigenpot.epsi,height=5.5cm,width=5.5cm} \qquad 
\epsfig{figure=theta.epsi,height=5.5cm,width=5.5cm}
\end{center}
\caption{(Left) Reduced eigen potentials (in ${\rm fm}^{-2}) $ $U_A (r)$ and
$U_R (r)$ as a function of the distance $r$ (in {\rm fm}). (Right) 
Local mixing angle $\theta(r)$  as a function of the distance $r$ 
(in {\rm fm}).}
\label{fig:theta}
\end{figure*}
At long distances we have the expansions
\begin{eqnarray}
\theta &=& \frac{ 2 \sqrt{2} R}{9 r} e^{-m r } (m^2 r^2 + 3 m r +3 ) +
\dots \\ U_A &=& - \frac{2 m^2 R}{3 r} e^{-m r} + \dots \\ U_R &=&
\frac{6}{r^2} + \frac{2R }{3r^3} ( m^2 r^2 + 6 m r + 6 ) e^{-m r} +
\dots
\end{eqnarray} 
whereas at short distances we have the behavior 
\begin{eqnarray}
\theta &=& \cos^{-1} \sqrt{\frac23}- \frac{r}{3\sqrt{2} R} +
\frac{r^2}{18 \sqrt{2} R^2} \nonumber \\ && - \frac{ (18 m^2 R^2 -5) r^3 }{324
\sqrt{2}R^3 }+ \dots \\ U_A &=& -\frac{4 R}{r^3}+
\frac{2}{r^2} - \frac{2}{3 r R} + \dots \\ U_R &=& \frac{8 R}{r^3} +
\frac{4}{r^2} + \frac{2 - 6 m^2 R^2}{3 R r} + \dots
\label{eq:rot_short}
\end{eqnarray} 
Note that in the locally rotated basis the mixing is related to the
derivative of the mixing angle, $\theta'$ which is small at all
distances (See Fig.~\ref{fig:theta}). Actually, at asymptotically
large distances we have
\begin{eqnarray}
u_A (r) \to u(r) \qquad u_R(r) \to w(r)
\end{eqnarray} 
If we neglect the mixing term in Eq.~(\ref{eq:sch_rotated}) the
equations decouple and, actually, there is no non-trivial solution for
the repulsive eigen-channel, since the energy is fixed
arbitrarily. Hence in the absence of mixing we have $ u_R=0$. At this
level of approximation we then get
\begin{eqnarray}
u(r) &=& \cos \theta (r) u_A(r) \\  
w(r) &=& \sin \theta (r) u_A(r) 
\end{eqnarray} 
In Fig.~\ref{fig:urotated} we show the solutions of the decoupled
equations compared to the exact ones. As we see, the difference in the
wave functions and hence the $D/S$ mixing is indeed small. Note that
this is {\it not} the same as to neglect the tensor force. The results
for the deuteron observables are presented in Table~\ref{tab:table1}. 
As we see, the quality of the zeroth $\eta$ approximation is rather good. 
\medskip

\begin{table*}
\caption{\label{tab:table1} Deuteron properties for the OPE potential.
We use the non-relativistic relation $ \gamma= \sqrt{ 2 \mu_{np} B} $
with $B=2.224575(9)$.  We compare the $\eta$-expansion at leading
order (LO), with standard perturbation theory at Next to leading order
(NLO) and the exact OPE result.}
\begin{ruledtabular}
\begin{tabular}{|c|c|c|c|c|c|c|}
\hline & $\gamma ({\rm fm}^{-1}$ & $\eta$ & $A_S ( {\rm fm}^{-1/2}) $
& $r_m ({\rm fm})$ & $Q_d ( {\rm fm}^2) $ & $P_D $ \\ \hline 
{\rm OPE}-$\eta$ (LO) & Input  &  0 & 0.8752 & 1.9423 &  0.1321 &  6\%\\ 
{\rm OPE}-pert (NLO) & Input & 0.051 & 0.7373 & 1.6429 & 0.4555 & 0  \\ 
{\rm OPE}-exact & Input & 0.02633 & 0.8681(1) & 1.9351(5) & 0.2762(1)
& 7.88(1)\% \\
NijmII & Input & 0.0253(2) & 0.8845(8) & 1.968(1) & 0.271(1) &
5.67(4)\% \\
Exp.  (non-rel.)  &  0.231605 &  0.0256(4)  & 0.8846(9)  & 1.971(6)  &
0.2859(3) & 5.67(4)\% \\
\end{tabular}
\end{ruledtabular}
\end{table*}

Actually, we can check {\it a posteriori} that the mixing is indeed
small for the zeroth order solutions. The inhomogeneous term at short
distances behaves as
\begin{eqnarray}
2 \theta'(r) u_A' (r) \to  - \frac{2}{3 \sqrt{2} R^2}
\left(\frac{r}{R}\right)^{-3/4} C_A \sin \left( 4 \sqrt{R/r} +
\alpha \right) \nonumber \\ 
\end{eqnarray} 
which compared to the remaining terms in Eq.~(\ref{eq:sch_rotated})
can indeed be considered small. Under these circumstances, the mixing
can then be included perturbatively, yielding
\begin{eqnarray}
u_R (r) = \int_0^\infty dr' G_R (r,r') \left[2 \theta'(r') u_A' (r') +
\theta''(r ') u_A(r') \right] \nonumber \\  
\label{eq:u_R-pert}
\end{eqnarray} 
with $G_R (r,r')$ the Green function of the homogeneous equation in
the repulsive eigen-channel, 
\begin{eqnarray}
G_R(r,r') &=& w_{\rm reg} (r) w_{\rm irreg} (r') \theta (r'-r) \nonumber \\ 
&+&  w_{\rm reg} (r') w_{\rm irreg} (r) \theta( r-r') 
\end{eqnarray} 
with Wronskian equal to unity and $w_{\rm reg} (r) $ and $w_{\rm irreg}
(r)$ the regular solution and irregular solutions at the origin
respectively. Asymptotically one has,
\begin{eqnarray} 
 w_{\rm reg} (r) &\to& C \left(\frac{r}{R}\right)^{3/4} e^{-4 \sqrt{2}
\sqrt{R/r}} \qquad \qquad ( r \to 0 ) \nonumber \\ w_{\rm reg} (r)
&\to& e^{+\gamma r} \left( 1 - \frac{3}{\gamma r} + \frac{3}{(\gamma
r)^2} \right) \qquad ( r \to \infty ) \nonumber \\ w_{\rm irreg} (r)
&\to& C \left(\frac{r}{R}\right)^{3/4} e^{+4 \sqrt{2} \sqrt{R/r}}
\qquad \qquad ( r \to 0 ) \nonumber \\ w_{\rm irreg} (r) &\to&
e^{-\gamma r} \left( 1 + \frac{3}{\gamma r} + \frac{3}{(\gamma r)^2}
\right) \qquad ( r \to \infty ) \nonumber \\
\label{eq:asym}
\end{eqnarray} 
To get in practice the coefficient $C$ we start with $C=1$ at short
distances and build the ratio to the asymptotic form at a sufficiently
large distance. With these conditions the solution $u_R (r) $ at large
distances behaves as
\begin{eqnarray} 
u_R (r) &\to& \eta e^{-\gamma r} \left( 1 + 
\frac{3}{\gamma r} + \frac{3}{(\gamma r)^2} \right)  \qquad ( r \to
\infty )  \nonumber\\
\end{eqnarray} 
with  
\begin{eqnarray}
\eta = \int_0^\infty dr w_{\rm reg} (r) \left[2 \theta'(r) u_A' (r) +
\theta''(r) u_A (r) \right]
\end{eqnarray} 
At short distances we get, from Eq.~(\ref{eq:u_R-pert}) and using the
asymptotic forms of Eq.~(\ref{eq:asym}) and Eq.~(\ref{eq:rot_short}), 
the result 
\begin{eqnarray} 
u_R (r) &\to& C C_A \left(\frac{r}{R}\right)^{7/4} \cos\left( 4 \sqrt{R/r} +
\alpha \right)
\end{eqnarray} 
in agreement with the leading short distance behavior of the full
solution for the combination $u (r) - \sqrt{2} w(r) $ (see
Sect.~\ref{sec:short}). The perturbative value for the asymptotic $D/S$
ratio we get is 
\begin{eqnarray}
\eta_{\rm pert} = 0.0261
\end{eqnarray} 
quite close to the OPE exact one, $ \eta_{\rm OPE}= 0.0263$.

\section{Long distance solutions} 

\begin{figure}[]
\begin{center}
\epsfig{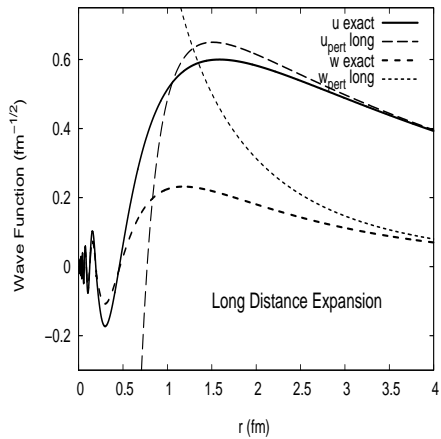}
\end{center}
\caption{The long distance expansion for the deuteron wave functions
matched. We take $\eta^{\rm OPE} =0.2633 $, and $A_S = 1$ 
for the norm of the numerical solutions. }
\label{fig:u+w_long.epsi}
\end{figure}

As a complement to the perturbative treatment of
Appendix~\ref{sec:pert} we analyze the bound solutions at long
distances.  The asymptotic deuteron wave functions for the OPE
potential can be written in the form
\begin{eqnarray}
u (r) &=& e^{-\gamma r} \left[ \sum_{k} F_{k} (r) e^{- k m
r} \right] \\ w(r) &=& \eta e^{-\gamma r} \left( 1 +
\frac{3}{\gamma r} + \frac{3}{(\gamma r)^2} \right) \left[ \sum_{k}
G_{k} (r) e^{- k m r} \right] \nonumber \\ 
\end{eqnarray} 
The first order solution can be evaluated analytically, yielding 
\begin{widetext} 
\begin{eqnarray} 
F_1(r) &=&  \frac{R\,e^{- m\,r} \left( m^2\,r^2 - 2\,\left( 1 + r\,\gamma  \right)  - 
       2\,m\,r\,\left( 1 + r\,\gamma  \right)  \right) \,\eta }{{\sqrt{2}}\,\,r^3\,
     {\gamma }^2} 
\nonumber \\ &+&
\frac{m^2\,R\,\left( 3\,{\sqrt{2}}\,m^2\,\eta  + 
       {\gamma }^2\,\left( 4 - 4\,{\sqrt{2}}\,\eta  \right)  \right) \,
     {\rm Ei}(-m\,r  )}{12\,{\gamma }^3} \nonumber \\ &+& 
  \frac{e^{2\,r\,\gamma }\,m^2\,R\,\left( -3\,{\sqrt{2}}\,m^2\,\eta  + 
       4\,{\gamma }^2\,\left( -1 + {\sqrt{2}}\,\eta  \right)  \right) \,
     {\rm Ei}(- m\,r  - 2\,r\,\gamma )}{12\,{\gamma }} \\ 
G_1(r) &=&  \frac{m^2\,R\,\left( -4\,{\gamma }^2\,\left(
       {\sqrt{2}} - 2\,\eta \right) + 3\,m^2\,\left( {\sqrt{2}} - \eta
       \right) \right) \, {\rm Ei}(- m\,r
      )}{12\,{\gamma }^3\,\eta } \nonumber \\ 
&+& \frac{R\, e^{- m\,r} \left( 3\,m^3\,r\,\left( {\sqrt{2}} - \eta \right) +
       m^2\,r\,\gamma \,\left( -3 + r\,\gamma \right) \,\left(
       {\sqrt{2}} - \eta \right) + 2\,{\gamma }^2 (1 + m r) \,\left( 2\,\eta +
       r\,\gamma \,\left( -{\sqrt{2}} + \eta \right) \right)  \right) }{2\,\,r\,
       {\gamma }^2\,\left( 3 + 3\,r\,\gamma + r^2\,{\gamma }^2 \right)
       \,\eta } \nonumber \\ 
&+& \frac{e^{2\,r\,\gamma
       }\,m^2\,R\,\left( 3 - 3\,r\,\gamma + r^2\,{\gamma }^2 \right)
       \, \left( 4\,{\gamma }^2\,\left( {\sqrt{2}} - 2\,\eta \right) -
       3\,m^2\,\left( {\sqrt{2}} - \eta \right) \right) \,
       {\rm Ei}(- m\,r  - 2\,r\,\gamma
       )}{12\,{\gamma }^3\, \left( 3 + 3\,r\,\gamma + r^2\,{\gamma }^2
       \right) \,\eta }  
\end{eqnarray} 
\end{widetext} 
The second order can also be evaluated but the expression is too long
to be presented here. In Fig.~\ref{fig:u+w_long.epsi} we present the
first order solutions compared to the exact ones. The
perturbative solutions of Appendix~\ref{sec:pert} are obtained from
the requirement that the S-wave $u$, becomes normalizable when
extended down to the origin. This can only happen in the $D/S$
asymptotic ratio, $\eta$ takes the value given by
Eq.~(\ref{eq:eta-pert}). This illustrates the fact that perturbation
theory can always be applied at long distances but fails at short
distances.

Note that here one treats the coupling constant $R$ and the mixing
parameter $\eta$ as independent variables.


\end{document}